\documentclass[sigconf,authorversion,nonacm]{acmart}

\PassOptionsToPackage{table,xcdraw}{xcolor} 
\usepackage{xcolor} 

\usepackage{dirtytalk} 
\usepackage{wrapfig}
\usepackage{subcaption}
\usepackage[bottom]{footmisc} 
\usepackage{booktabs}
\usepackage{graphicx}
\usepackage{multirow}
\usepackage{pifont}
\usepackage{colortbl}

\definecolor{myBlue}{RGB}{0, 122, 255}
\definecolor{myBlack}{RGB}{0, 0, 0}
\definecolor{myRed}{RGB}{190,55,74}

\newcommand{\sayit}[1]{{\textit{\say{#1}}}}

\AtBeginDocument{%
  }

\copyrightyear{2025}
\acmYear{2025}
\setcopyright{cc}
\setcctype{by}
\acmConference[CHI '25]{CHI Conference on Human Factors in Computing Systems}{April 26-May 1, 2025}{Yokohama, Japan}
\acmBooktitle{CHI Conference on Human Factors in Computing Systems (CHI '25), April 26-May 1, 2025, Yokohama, Japan}\acmDOI{10.1145/3706598.3713421}
\acmISBN{979-8-4007-1394-1/25/04}

\begin{document}

\title[Accessibility for Whom? ]{Accessibility for Whom? \\ Perceptions of Sidewalk Barriers Across Disability Groups and Implications for Designing Personalized Maps}


\author{Chu Li}
\orcid{0009-0003-7612-6224}
\affiliation{%
    \institution{Allen School of Computer Science}
  \institution{University of Washington, USA}
  \country{}
}
\email{chuchuli@cs.washington.edu}

\author{Rock Yuren Pang}
\orcid{0000-0001-8613-498X}
\affiliation{%
    \institution{Allen School of Computer Science}
  \institution{University of Washington, USA}
  \country{}
}
\email{ypang2@cs.washington.edu}

\author{Delphine Labbé}
\orcid{0000-0002-3730-4310}
\affiliation{%
    \institution{Disability and Human Development}
  \institution{University of Illinois at Chicago, USA}
  \country{}
}
\email{dlabbe@uic.edu}

\author{Yochai Eisenberg}
\orcid{0000-0001-5598-771X}
\affiliation{%
    \institution{Disability and Human Development}
  \institution{University of Illinois at Chicago, USA}
  \country{}
}
\email{yeisen2@uic.edu}

\author{Maryam Hosseini}
\orcid{0000-0002-4088-810X}
\affiliation{%
    \institution{City and Regional Planning}
  \institution{UC Berkeley, USA}
  \country{}
}
\email{maryamh@berkeley.edu}

\author{Jon E. Froehlich}
\orcid{0000-0001-8291-3353}
\affiliation{%
    \institution{Allen School of Computer Science}
  \institution{University of Washington, USA}
  \country{}
}
\email{jonf@cs.washington.edu}

\renewcommand{\shortauthors}{Li et al.}

\begin{abstract}
Today's mapping tools fail to address the varied experiences of different mobility device users. This paper presents a large-scale online survey exploring how five mobility groups---users of canes, walkers, mobility scooters, manual wheelchairs, and motorized wheelchairs---perceive sidewalk barriers and differences therein. Using 52 sidewalk barrier images, respondents evaluated their confidence in navigating each scenario. Our findings (\textit{N=}190) reveal variations in barrier perceptions across groups, while also identifying shared concerns. To further demonstrate the value of this data, we showcase its use in two custom prototypes: a visual analytics tool and a personalized routing tool. Our survey findings and open dataset advance work in accessibility-focused maps, routing algorithms, and urban planning.

\end{abstract}


\begin{teaserfigure}
  \includegraphics[width=\textwidth]{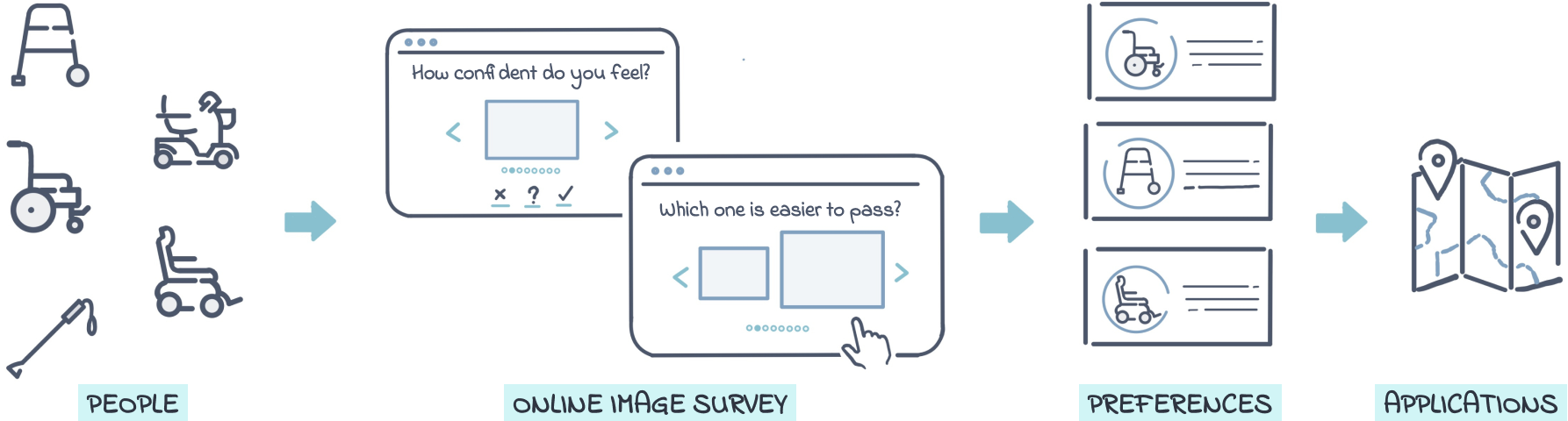}
  \caption{We present a large-scale online image survey gathering perceptions of sidewalk barriers from five mobility groups: users of walking canes, walkers, mobility scooters, manual wheelchairs, and motorized wheelchairs. Findings were used to generate user profiles that informed the design of personalized accessibility maps and routing tools.}
  \Description{This figure shows our pipeline for large-scale online image survey gathering perceptions of sidewalk barriers from five mobility groups: users of walking canes, walkers, mobility scooters, manual wheelchairs, and motorized wheelchairs. Findings were used to generate user profiles that informed the design of personalized accessibility maps and routing tools.}
  \label{fig:teaser}
\end{teaserfigure}


\maketitle

\section{Introduction}

\begin{figure*}
    \centering
    \includegraphics[width=1\linewidth]{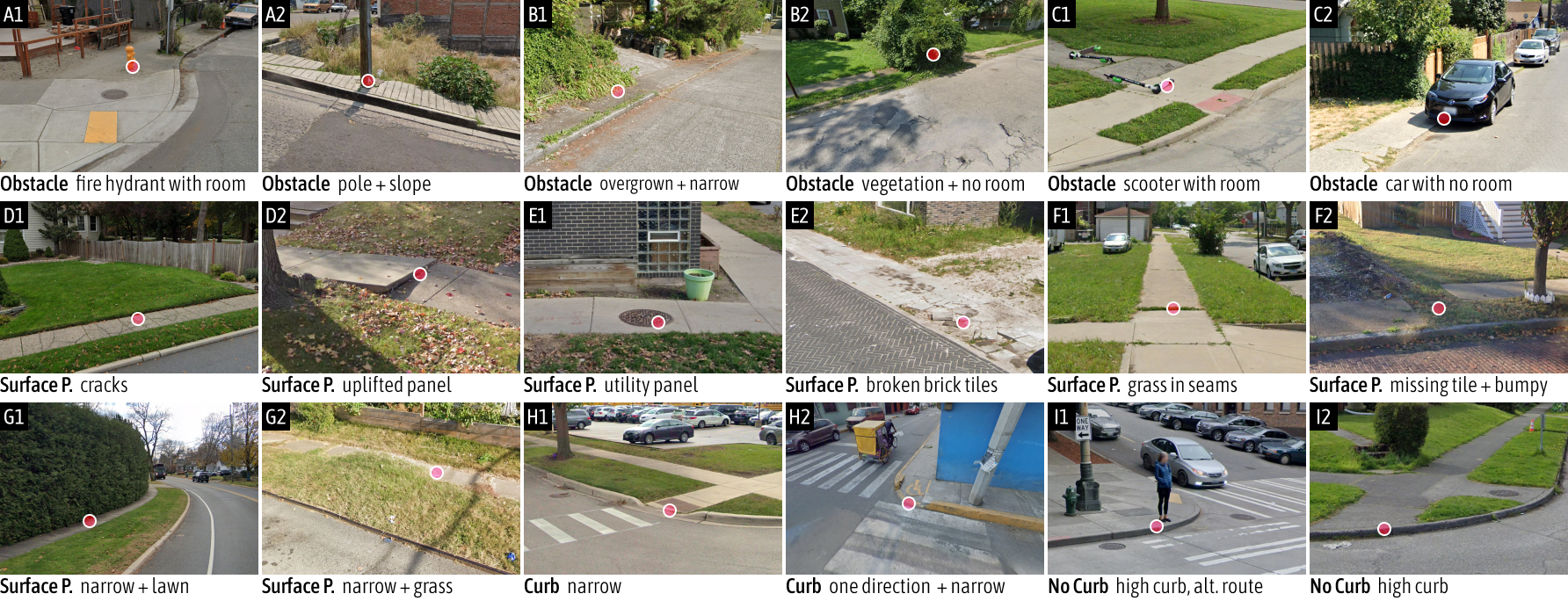}
    \caption{Our final image dataset consists of 52 images across nine categories, including \textit{fire hydrants + poles}, \textit{overgrown vegetation}, \textit{parked bikes/cars/scooters}, \textit{cracks/height differences}, \textit{bricks/cobblestone + utility panels }, \textit{sand/gravel + grass}, \textit{narrow}, \textit{curb ramp}, \textit{missing curb ramp},  Above, we show two sample images from each category.}
    \Description{
    This figure shows our final image dataset composed of 52 images across nine categories, including poles/fire hydrants, vegetation, parked bikes/cars, cracks/broken surfaces, bricks/cobblestone, sand/gravel/grass, narrow, curb ramp, missing curb ramp, In this figure, we show two sample images from each of the nine categories.
    }
    \label{fig:image-dataset-grid}
\end{figure*}

In 2022, over 18 million U.S. adults reported having a mobility-related disability, with nearly half (49.3\%) using an assistive aid such as a cane, crutches, walker, scooter or wheelchair~\cite{firestine_travel_2024}. Mobility aid users confront an array of environmental barriers in their everyday travel, such as missing curb ramps, uneven sidewalks, and major obstacles on the sidewalk like impassable street furniture or overgrown vegetation~\cite{meyers_barriers_2002, ding_design_2007,rosenberg_outdoor_2013}. These challenges can be mitigated by offering directions that avoid barriers and guide mobility aid users to destinations safely, accurately, and efficiently. However, current navigation systems (\textit{e.g.,} Google Maps) and commercial analytical services (\textit{e.g.,} Walk Score~\cite{walk_score_walk_2007}) fail to account for the unique requirements and preferences of people with mobility disabilities. Several research projects have developed routing and mapping systems that incorporate mobility disabilities, such as \textit{MAGUS}~\cite{matthews_modelling_2003}, \textit{U-Access}~\cite{sobek_u-access_2006}, \textit{AccessScore}~\cite{li_interactively_2018}, and \textit{AccessMap}~\cite{bolten_accessmap_2019}.
Though promising, prior work often overlooks the heterogeneity among users of different devices and focuses predominantly on wheelchair  users~\cite{kasemsuppakorn_personalised_2009,saha_project_2019,kasemsuppakorn_understanding_2015,matthews_modelling_2003}. However, a larger percentage of users gain mobility from canes, crutches, or walkers~\cite{firestine_travel_2024}. This contrast underscores the need to incorporate a wider range of mobility aid users into our mapping tools and to better characterize the unique challenges each group faces ~\cite{shoemaker_development_2010}. For example, a missing curb ramp at an intersection may pose a significant barrier for wheelchair and mobility scooter users but is less challenging for those using walking canes or walkers. 

To examine perceptions of sidewalk barriers across disability groups and to inform the design of future personalized, disability-aware maps, we developed a large-scale online image survey for five mobility groups: walking cane, walker, mobility scooter, manual wheelchair, and motorized wheelchair users. The survey featured a curated set of 52 sidewalk barrier images collected through an online crowdsourcing platform called Project Sidewalk~\cite{saha_project_2019}, the images include nine barrier categories: fire hydrant \& pole, tree \& vegetation, parked bikes/scooter/cars, height difference \& sidewalk cracks, manholes \& brick/cobblestone, grass \& sand/gravel, narrow and (missing) curb ramps (\autoref{fig:image-dataset-grid}). We asked respondents to evaluate their confidence in passing the scenarios from the images while using their respective mobility aid(s). The survey used a combination of rating, ranking, and adaptive pairwise comparison as well as open-ended text questions.

Our findings (\textit{N=}190) suggest that walking cane users were more likely to perceive sidewalk issues as passable, while mobility scooter users were more likely to perceive sidewalk issues as challenges. Although each group faces unique barriers, high-severity obstacles, surface problems, and missing curb ramps impose significant restrictions for all users. Interestingly, while the groups generally agree on the passability of low and high severity issues, their perceptions diverge for mid-severity issues. Notably, wheeled mobility users show higher sensitivity to missing curb ramps compared to walking cane and walker users. Despite variations in assessing individual sidewalk issues, users across all mobility aid types demonstrate consistent judgments when ranking images within each category from most to least passable.

To demonstrate the value of this data to HCI, accessibility, and urban planning researchers, we created two example applications. First, we synthesized user preferences from our findings, to create \textit{interactive accessibility rating maps} based on each user group’s perceived passability. These maps reveal similar yet distinct patterns across different mobility device groups, providing nuanced views of city-wide accessibility levels. Second, we created a \textit{disability-aware routing prototype} based on OSMnx~\cite{boeing_osmnx_2017} to generate personalized, optimal paths for each mobility group. These applications showcase the potential of our data to inform those with disabilities about residential and social choices, provide personalized route planning strategies, and develop analytical tools that identify obstacles and assess the impact of their removal.

\begin{table*}[t]
\centering
\renewcommand{\arraystretch}{1.25}
\resizebox{\linewidth}{!}{%
\begin{tabular}{l|lll|llll|l|l}
 &
  \multicolumn{3}{c|}{\textbf{Obstacle}} &
  \multicolumn{4}{c|}{\textbf{Surface Problems}} &
  \textbf{Curb Ramps} &
  \textbf{Missing Curb Ramps} \\ 
  \hline
\rowcolor[HTML]{F3F3F3}
\textbf{Subcategories} &
  \begin{tabular}[c]{@{}l@{}}Fire hydrant \\ + pole\end{tabular} &
  \begin{tabular}[c]{@{}l@{}}Overgrown \\ vegetation\end{tabular} &
  \begin{tabular}[c]{@{}l@{}}Parked cars, \\ bikes, scooters\end{tabular} &
  \begin{tabular}[c]{@{}l@{}}Cracks \\ + height \\ differences\end{tabular} &
  \begin{tabular}[c]{@{}l@{}}Brick/\\ cobblestone, \\ utility panels\end{tabular} &
  \begin{tabular}[c]{@{}l@{}}Sand/gravel \\ + grass\end{tabular} &
  Narrow &
  Curb Ramps &
  Missing Curb Ramps \\ 
\textbf{Severities} &
  \begin{tabular}[c]{@{}l@{}}2 low, \\ 2 mid, \\ 2 high\end{tabular} &
  \begin{tabular}[c]{@{}l@{}}2 low, \\ 2 mid, \\ 2 high\end{tabular} &
  \begin{tabular}[c]{@{}l@{}}2 low, \\ 2 mid, \\ 2 high\end{tabular} &
  \begin{tabular}[c]{@{}l@{}}2 low, \\ 2 mid, \\ 2 high\end{tabular} &
  \begin{tabular}[c]{@{}l@{}}2 low, \\ 2 mid, \\ 2 high\end{tabular} &
  \begin{tabular}[c]{@{}l@{}}2 low, \\ 2 mid, \\ 2 high\end{tabular} &
  \begin{tabular}[c]{@{}l@{}}2 low, \\ 2 mid, \\ 2 high\end{tabular} &
  \begin{tabular}[c]{@{}l@{}}2 low, \\ 2 mid, \\ 2 high\end{tabular} &
  \begin{tabular}[c]{@{}l@{}}2 low, \\ 2 mid, \\ 2 high\end{tabular} \\
\end{tabular}%
}
\vspace{1em}
\caption{We organized the final image dataset hierarchically across four top-level categories: \textit{obstacles}, \textit{surface problems}, \textit{curb ramps}, and \textit{missing curb ramps} as well as nine subcategories. For each subcategory, we selected two low, mid, and high severity images (as drawn from Project Sidewalk ratings). One exception was the \textit{narrow} subcategory for surface problems, which had two low and mid subcategories only.}
\label{tab:label-categories}
\end{table*}

In summary, our contributions are threefold: 
(1) we present \textit{results from a large-scale survey of people with diverse mobility aids}, providing insights into how specific mobility aids shape people’s perceptions of the built environment;
(2) we demonstrate \textit{how to apply these findings} to generate accessibility rating maps and enhance personalized routing algorithms;
and (3) we contribute an open-source dataset and our analysis code\footnote{\href{https://github.com/makeabilitylab/accessibility-for-whom}{https://github.com/makeabilitylab/accessibility-for-whom}}, enabling researchers and developers to leverage this work to further promote accessibility solutions for mobility aid users. Our work advances HCI/accessibility research and urban planning by complementing existing sidewalk datasets with diverse perspectives from multiple mobility aid users. Our dataset and findings enable the development of more accurate, tailored routing algorithms for people with different mobility needs, while providing urban planners and policymakers with crucial data to prioritize and target accessibility improvements and renovations.

\section{Related Work}
Our work draws on and contributes to research in mobility aids and the built environment, online image-based survey for urban assessment, personalized routing applications and accessibility maps.

\subsection{Mobility Aids and the Built Environment}
People who use mobility aids (\textit{e.g.,} canes, walkers, mobility scooters, manual wheelchairs and motorized wheelchairs) face significant challenges navigating their communities.
Studies have repeatedly found that sidewalk conditions can significantly impede mobility among these users~\cite{bigonnesse_role_2018,fomiatti_experience_2014,f_bromley_city_2007,rosenberg_outdoor_2013, harris_physical_2015,korotchenko_power_2014}. 
In a review of the physical environment's role in mobility, \citet{bigonnesse_role_2018} summarized factors affecting mobility aid users, including uneven or narrow sidewalks (\textit{e.g.,}~\cite{fomiatti_experience_2014,f_bromley_city_2007}), rough pavements (\textit{e.g.,}~\cite{fomiatti_experience_2014,f_bromley_city_2007}), absent or poorly designed curb ramps (\textit{e.g.,}~\cite{rosenberg_outdoor_2013, f_bromley_city_2007, korotchenko_power_2014}), lack of crosswalks (\textit{e.g.,}~\cite{harris_physical_2015}), and various temporary obstacles (\textit{e.g.,}~\cite{harris_physical_2015}).

Though most research on mobility disability and the built environment has focused on wheelchair users~\cite{bigonnesse_role_2018}, mobility challenges are not experienced uniformly across different user populations~\cite{prescott_factors_2020, bigonnesse_role_2018}. 
For example, crutch users could overcome a specific physical barrier (such as two stairs down to a street), whereas motorized wheelchair users could not (without a ramp)~\cite{bigonnesse_role_2018}. 
Such variability demonstrates how person-environment interaction can differ based on mobility aids and environmental factors~\cite{sakakibara_rasch_2018,smith_review_2016}.
Further, mobility aids such as canes, crutches, or walkers are more commonly used than wheelchairs in the U.S.~\cite{taylor_americans_2014, firestine_travel_2024}: in 2022, approximately 4.7 million adults used a cane, crutches, or a walker, compared to 1.7 million who used a wheelchair~\cite{firestine_travel_2024}.
This underscores the importance of considering a diverse range of mobility aid users in urban accessibility research.
For example, \citet{prescott_factors_2020} explored the daily path areas of users of manual wheelchairs, motorized wheelchairs, scooters, walkers, canes, and crutches and found that the type of mobility device had a strong association with users' daily path area size.
Our study aims to further advance knowledge of how different mobility aid users perceive sidewalk barriers, with a more inclusive understanding of urban accessibility.

\begin{figure*}
    \centering
    \includegraphics[width=1\linewidth]{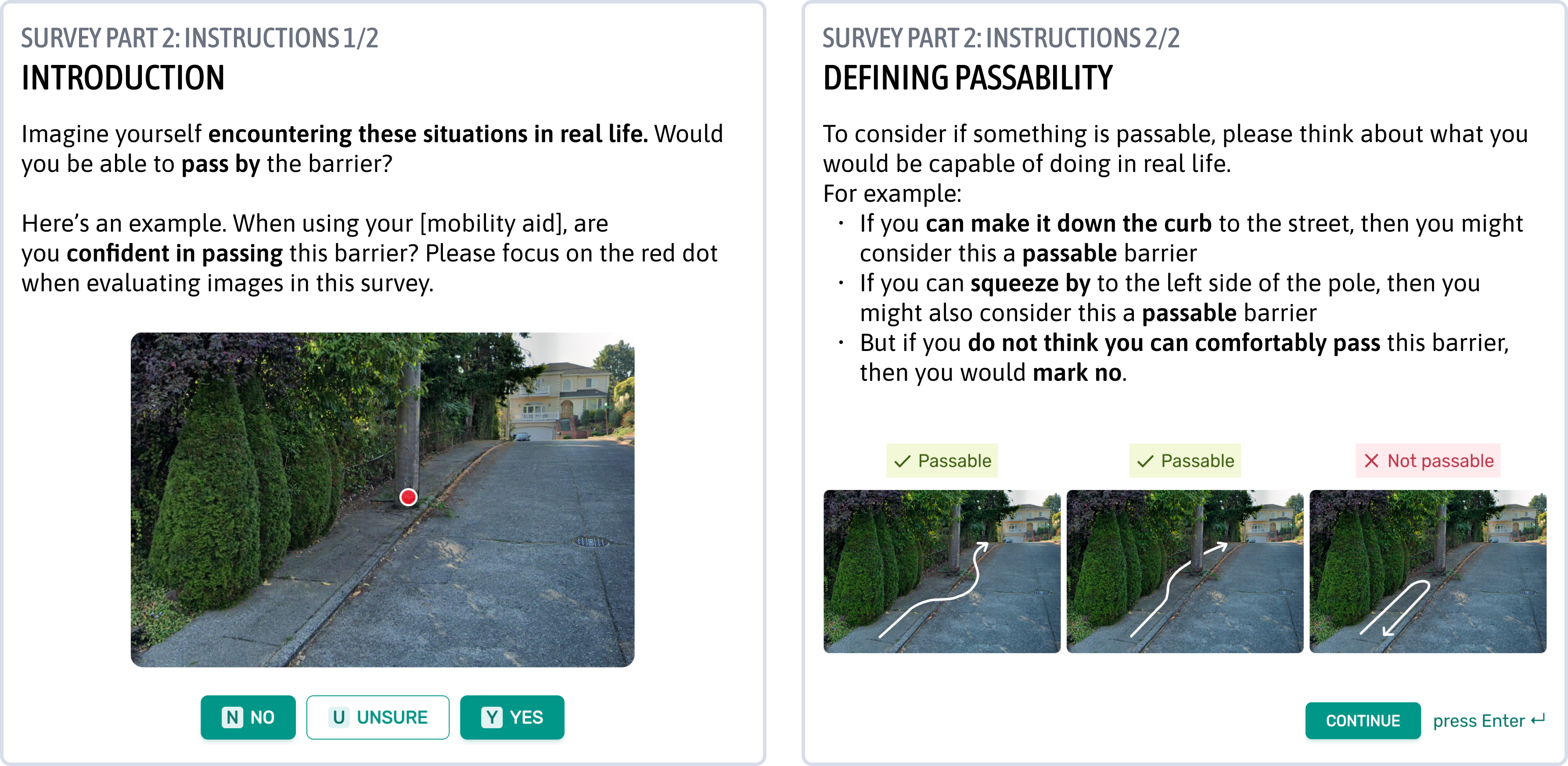}
    \caption{Survey Part 2.1 showed all 52 images and asked participants to rate their passability based on their lived experience and use of their mobility aid. Above is the interactive tutorial we showed at the beginning of this part.}
    \Description{This figure shows a screenshot from the online survey. In survey part 2.1, participants were presented with 52 images and were asked to rate their passibility based on their lived experience and use of their mobility aid. The screenshot shows the interactive tutorial shown before this section.}
    \label{fig:survey-part2-instructions}
\end{figure*}

\subsection{Online Image-Based Survey for Urban Assessment}
Sidewalk barriers hinder individuals with mobility impairments not just by preventing particular travel paths but also by reducing confidence in self-navigating and decreasing one's willingness to travel to areas that might be physically challenging or unsafe~\cite{vasudevan_exploration_2016,clarke_mobility_2008}.
Prior work in this area traditionally uses three main study methods: in-person interviews (\textit{e.g}.~\cite{rosenberg_outdoor_2013,castrodale_mobilizing_2018}), GPS-based activity studies (\textit{e.g.,}~\cite{prescott_exploration_2021, prescott_factors_2020,rosenberg_outdoor_2013}), and online-questionnaires (\textit{e.g.,}~\cite{carlson_wheelchair_2002}). 
In-person interviews, while providing detailed and nuanced information, are limited by small sample sizes~\cite{rosenberg_outdoor_2013}. GPS-based activity studies involve tracking mobility aids user activity over a period of time, offering insights into movement patterns and activity space; however, these studies are constrained by geographical location~\cite{prescott_exploration_2021}. In contrast, online questionnaires can reach much larger populations and cover broader geographical regions, but they often yield high-level information that lacks the depth and nuance of the other approaches~\cite{carlson_wheelchair_2002}.
Our study aims to strike a balance between these approaches, capturing nuanced perspectives of mobility aid users about the built environment while maintaining a sufficiently large enough sample size for robust statistical analysis. 
Building on~\citet{bigonnesse_role_2018}'s work, we explore not only the types of factors considered to be barriers, but the \textit{intensity} of these barriers and their differential impacts.

Visual assessment of environmental features has long been employed by researchers across diverse fields, including human well-being~\cite{humpel_environmental_2002}, ecosystem sustainability~\cite{gobster_shared_2007}, and public policy~\cite{dobbie_public_2013}. 
These studies examine the relationship between images and the reactions they provoke in respondents or compare differences in reactions between groups.
Over the past decade, online visual preference surveys have gained popularity (\textit{e.g.,}~\cite{evans-cowley_streetseen_2014, salesses_collaborative_2013, goodspeed_research_2017}), where respondents are asked to make pairwise comparisons between randomly selected images.
Using this approach has two advantages: it adheres to the law of comparative judgment~\cite{thurstone_law_2017} by allowing respondents to make direct comparisons, and it prevents inter-rater inconsistency possible with scale ratings~\cite{goodspeed_research_2017}.
Additionally, online surveys generally offer advantages of increased sample sizes, reduced costs, and greater flexibility~\cite{wherrett_issues_1999}.
For people with disabilities, online surveys can be particularly beneficial. They help reach hidden or difficult-to-access populations~\cite{cook_challenges_2007,wright_researching_2005} and are believed to encourage more honest answers to sensitive questions~\cite{eckhardt_research_2007} by providing a higher level of anonymity and confidentiality~\cite{cook_challenges_2007, wright_researching_2005}.

\begin{figure*}
    \centering
    \includegraphics[width=1\linewidth]{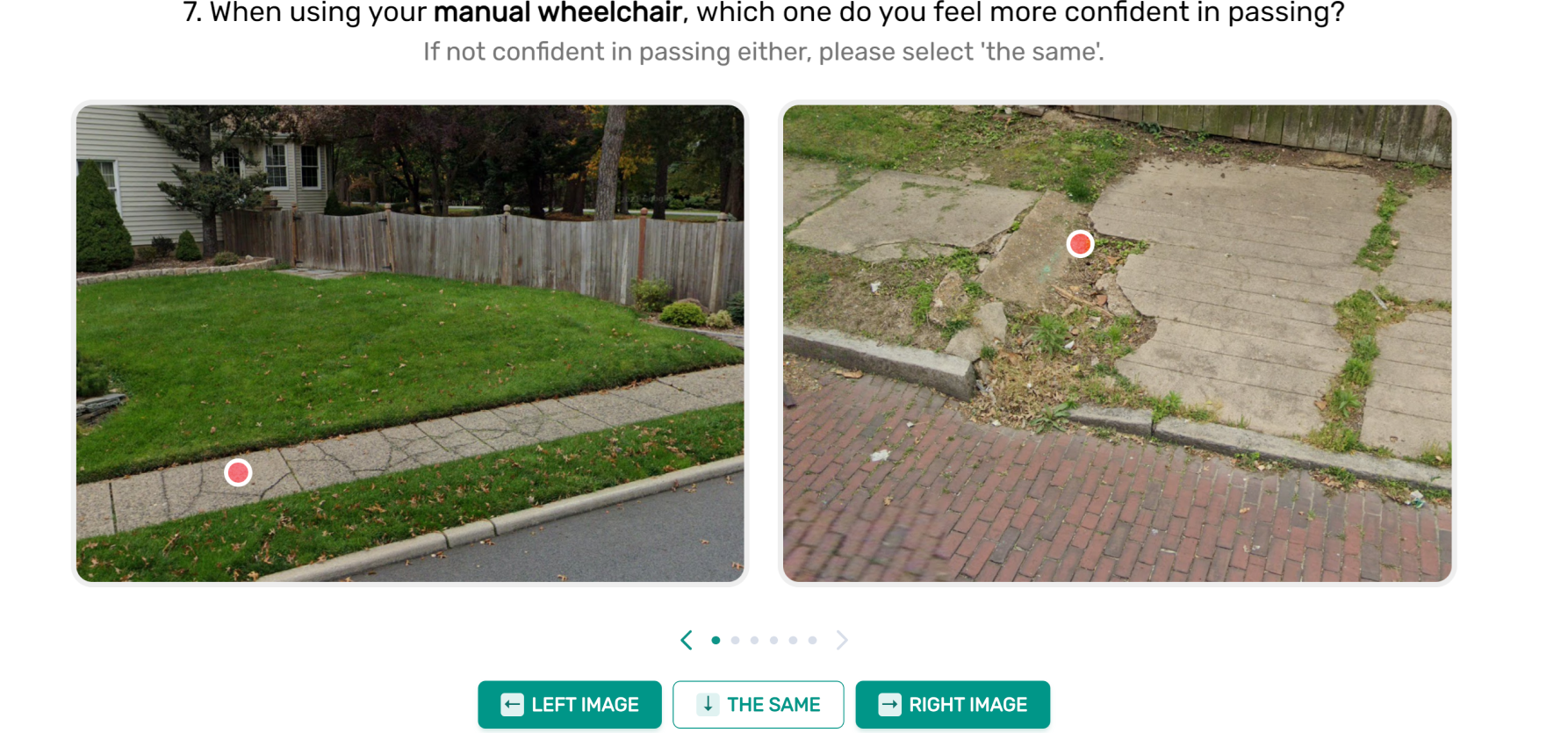}
    \caption{In survey Part 2.2, participants were asked to perform a series of pairwise comparisons based on their 2.1 responses.}
    \Description{This figure shows a screenshot from the online survey. In Survey Part 2.2, participants were asked to perform a series of pairwise comparisons based on their 2.1 responses.}
    \label{fig:survey-part2b-pairwise}
\end{figure*}

\subsection{Personalized Routing Applications and Accessibility Maps}
Navigation challenges faced by mobility aid users can be mitigated through the provision of routes and directions that guide them to destinations safely, accurately, and efficiently~\cite{kasemsuppakorn_understanding_2015}. However, current commercial routing applications (\textit{e.g.}, \textit{Google Maps}) do not provide sufficient guidance for mobility aid users.
To address this gap, significant research has focused on routing systems for this population over the past two decades~\cite{barczyszyn_collaborative_2018, karimanzira_application_2006, matthews_modelling_2003, kasemsuppakorn_understanding_2015, volkel_routecheckr_2008, holone_people_2008, wheeler_personalized_2020, gharebaghi_user-specific_2021, ding_design_2007}.
One early, well-known prototype system is \textit{MAGUS}~\cite{matthews_modelling_2003}, which computes optimal routes for wheelchair users based on shortest distance, minimum barriers, fewest slopes, and limits on road crossings and challenging surfaces.
\textit{U-Access}~\cite{sobek_u-access_2006} provides the shortest route for people with three accessibility levels: unaided mobility, aided mobility (using crutch, cane, or walker), and wheelchair users.
However, U-Access only considers distance and ignores other
important factors for mobility aid users~\cite{barczyszyn_collaborative_2018}.
A series of projects by Kasemsuppakorn \textit{et al}.~\cite{kasemsuppakorn_personalised_2009, kasemsuppakorn_understanding_2015} attempted to create personalized routes for wheelchair users using fuzzy logic and \textit{Analytic Hierarchy Process} (AHP).

While influential, many personalized routing prototypes face limited adoption due to a scarcity of accessibility data for the built environment. 
Geo-crowdsourcing~\cite{karimi_personalized_2014}, a.k.a. volunteered geographic information (VGI)~\cite{goodchild_citizens_2007}, has emerged as an effective solution~\cite{karimi_personalized_2014, wheeler_personalized_2020}.
In this approach, users annotate maps with specific criteria or share personal experiences of locations, typically using web applications based on Google Maps or \textit{OpenStreetMap} (OSM)~\cite{karimi_personalized_2014}.
Examples include \textit{Wheelmap}~\cite{mobasheri_wheelmap_2017}, \textit{CAP4Access}~\cite{cap4access_cap4access_2014}, \textit{AXS Map}~\cite{axs_map_axs_2012}, and \textit{Project Sidewalk}~\cite{saha_project_2019}.
Recent research demonstrated the potential of using crowdsourced geodata for personalized routing~\cite{goldberg_interactive_2016, bolten_accessmap_2019,menkens_easywheel_2011, neis_measuring_2015}.
For example, \textit{EasyWheel}~\cite{menkens_easywheel_2011}, a mobile social navigation system based on OSM, provides wheelchair users with optimized routing, accessibility information for points of interest, and a social community for reporting barriers. 
\textit{AccessMap}~\cite{bolten_accessmap_2019} offers routing information tailored to users of canes, manual wheelchairs, or powered wheelchairs, calculating routes based on OSM data that includes slope, curbs, stairs and landmarks. 
Our work builds on the above by gathering perceptions of sidewalk obstacles from different mobility aid users to create generalizable profiles based on mobility aid type. We envision that these profiles can provide starting points in tools like Google Maps for personalized routing but can be further customized by the end user to specify additional needs (\textit{e.g.}, ability to navigate hills, \textit{etc.})

Beyond routing applications, our study data can contribute to modeling and visualizing higher-level abstractions of accessibility. 
Similar to \textit{AccessScore}~\cite{li_interactively_2018}, data from our survey can provide personalizable and interactive visual analytics of city-wide accessibility. By identifying both differences between mobility groups and common barriers within groups, we can develop analytical tools to prioritize barriers and assess the impact of their mitigation or removal, potentially benefiting the broadest range of mobility group users. Incorporating perceptions of passibility into urban planning processes provides a new dimension for urban planners' toolkits, which are often narrowly focused on compliance with ADA standards.

\section{Method}

To study how people who use mobility aids perceive sidewalk barriers, we created an image-based online survey. 
Participants were shown a curated set of images that contain sidewalk barriers and asked to respond accordingly based on their lived experience and their mobility aid usage. 
We aim to address three overarching research questions (RQs):
\begin{itemize}
    \item RQ1: How do people with different mobility aids perceive mobility barriers?
    \item RQ2: What are the key similarities and differences between mobility aid groups?
    \item RQ3: What types of barriers are the most severe and why?
\end{itemize}

Below, we describe our study method, including the iterative survey design and development, participant recruitment, and our data and analysis. 

\subsection{Study Method}
We begin by outlining our five primary mobility aid groups, followed by an overview of our sidewalk image dataset and the survey.

\subsubsection{User Groups}
Many different types of sensory~\cite{giudice_blind_2008,raina_relationship_2004}, cognitive~\cite{pillette_systematic_2023,buchman_cognitive_2011}, and physical disabilities~\cite{groessl_physical_2019,garg_associations_2016} can impact mobility. 
We specifically focus on people with ambulatory disabilities that require a mobility aid, including \textit{walking canes}, \textit{walkers}, \textit{mobility scooters}, \textit{manual wheelchairs}, and \textit{motorized wheelchairs}.
These categories were informed by the \textit{National Household Travel Survey} (NHTS)~\cite{us_department_of_transporation_national_2022}, \textit{National Survey on Health and Disability} (NSHD)~\cite{the_university_of_kansas_national_2018}, and \textit{Canadian Survey on Disability}~\cite{government_of_canada_canadian_2022}, as well as insights from our multi-disciplinary research team. 
We included the mobility devices that appeared consistently across all three surveys~\cite{us_department_of_transporation_national_2022, the_university_of_kansas_national_2018, government_of_canada_canadian_2022}, except for \textit{crutches}, which we excluded as a main category due to their temporary nature~\cite{manocha_injuries_2021}.
We also excluded several other aids as main categories: \textit{white canes}~\cite{us_department_of_transporation_national_2022, the_university_of_kansas_national_2018}, as they are typically used by blind and low-vision individuals and our survey relied on visual examination of images; \textit{artificial limbs} or \textit{prosthetics}~\cite{the_university_of_kansas_national_2018, government_of_canada_canadian_2022}, as they are highly customized to individuals and provide fewer insights when studying mobility aid users as a group; and \textit{service animals}~\cite{us_department_of_transporation_national_2022, the_university_of_kansas_national_2018}, as they are not \textit{device aids}. 
\autoref{tab:survey-groups} provides a comprehensive list of aids mentioned in the three surveys, accompanied by a rationale for each aid that was excluded from our study as a user group.
To acknowledge the diversity of mobility devices, we include an \textit{other} category where users can specify alternative mobility aids not covered by the main categories.

\begin{figure*}
    \centering
    \includegraphics[width=0.75\linewidth]{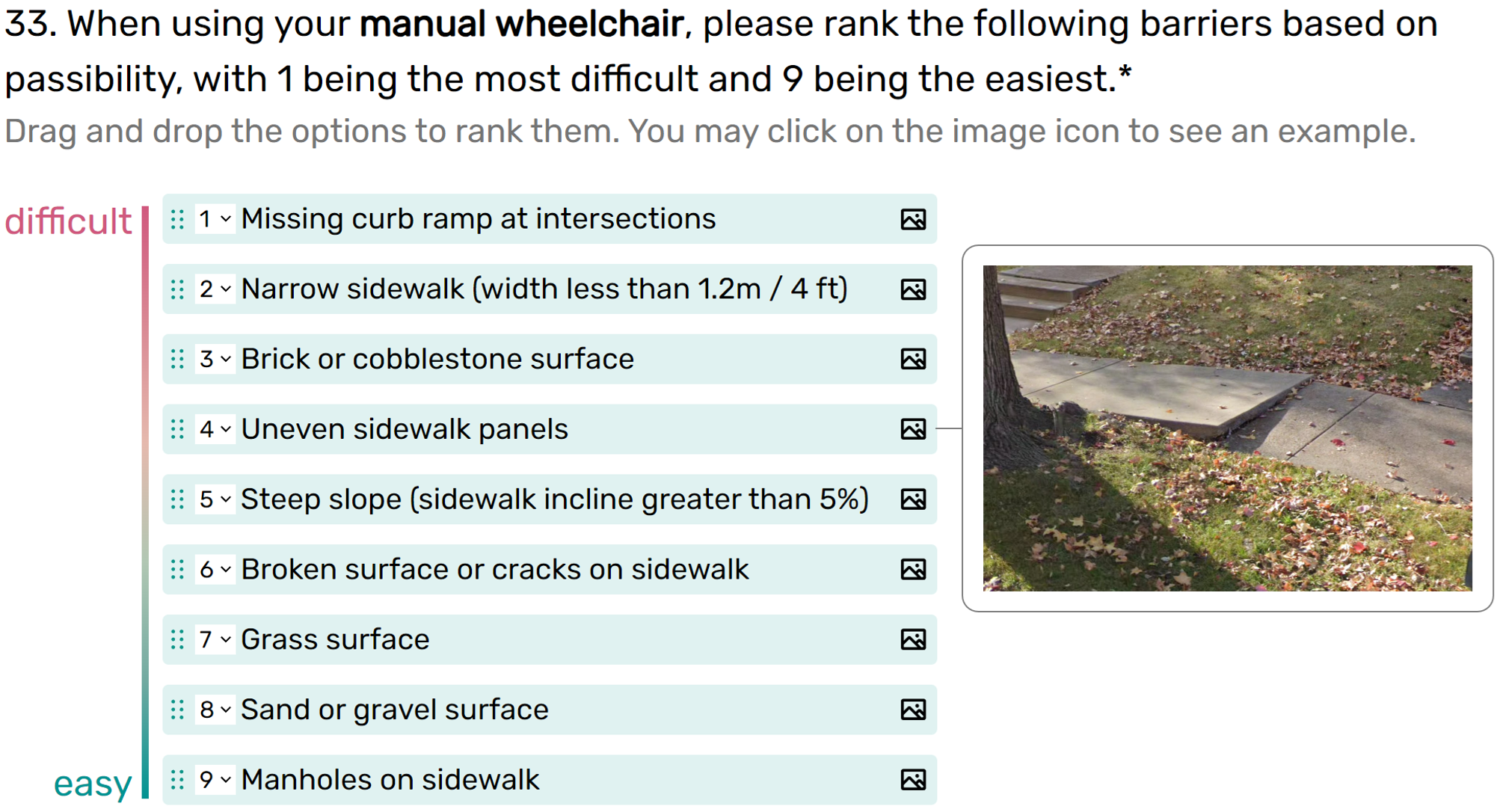}
    \caption{In Part 3, the survey asked participants to rank order nine types of sidewalk barriers, including: \textit{missing curb ramps}, \textit{narrow sidewalks}, \textit{brick/cobblestone surfaces}, \textit{uneven sidewalk panels}, \textit{steep slopes}, \textit{broken surfaces/cracks}, \textit{grass surfaces}, \textit{sand/gravel surfaces}, and \textit{manholes on sidewalks}. To elicit the most accurate responses, participants could click on the image icon to see an example image for each barrier.}
    \Description{This figure shows a screenshot from the online survey. In Part 3, the participants were asked to rank order nine types of sidewalk barriers, including: missing curb ramps, narrow sidewalks, brick/cobblestone surfaces, uneven sidewalk panels, steep slopes, broken surface/cracks, grass surfaces, sand/gravel surfaces and manholes on sidewalks. To ground responses, participants could click on the image icon to see an example image for each  barrier.}
    \label{fig:survey-part3-ranking}
\end{figure*}

\subsubsection{Sidewalk Image Dataset}
Our survey uses images as the primary stimulus: we show participants example sidewalk images and ask them to respond given their lived experience and specific mobility aid usage. 
The research team selected these images from \textit{Project Sidewalk}~\cite{saha_project_2019}---an open-source web tool where users virtually find, label, and rate sidewalk conditions through interactive streetscape imagery.
Project Sidewalk is deployed in 21 cities across eight countries, amassing over one million labeled sidewalk images. 
As the built environment can differ by country and geographic context (\textit{e.g.,} rural \textit{vs.} urban), our initial focus is on studying North American infrastructure. Thus, our images derive primarily from North American cities: Seattle, WA; Oradell, NJ; Chicago, IL; Columbus, OH; and Mexico City, Mexico.

To select and curate our image dataset, we used Project Sidewalk's \textit{Image Gallery} tool~\cite{duan_sidewalk_2021}, which provides an interactive gallery of all labeled sidewalk images filterable based on seven high-level sidewalk feature and barrier categories (\textit{e.g.,} curb ramps, surface problems, obstacles), ~40 tag categories (\textit{e.g.,} uplifts, cracks, cobblestone), and a five-point severity scale. Through an iterative process of selection and discussion across three research members, we finalized a dataset of 52 images covering four major label types and nine tag categories---see \autoref{fig:image-dataset-grid} and ~\autoref{tab:label-categories}. To  pinpoint key issues, we consolidated the five-point severity scale into a three-point scale: high, medium, and low. For each tag category, we selected two images per severity level, as shown in \autoref{tab:label-categories}.

Our overarching goal was to curate a dataset that showcased a variety of common sidewalk accessibility problems of varying severities. To allow others to build on our research, our dataset is available on Github\footnote{\href{https://github.com/makeabilitylab/accessibility-for-whom}{https://github.com/makeabilitylab/accessibility-for-whom}}.

\begin{figure*}
    \centering
    \includegraphics[width=1\linewidth]{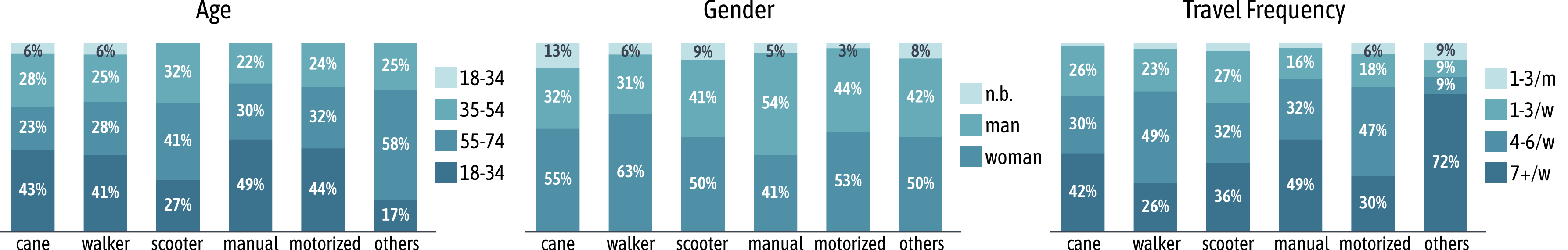}
    \caption{Participant demographics (\textit{N=}144) broken down by mobility aid, including age, gender, and travel frequency. \textit{n.b.} is non-binary; \textit{1-3/m} is 1-3 travels per month or less,  \textit{1-3/w} is 1-3 travels per week, \textit{4-6/w} is 4-6 travels per week, \textit{7+/w} is 7 travels per week or more.}
    \Description{This is a data visualization of bar plots showing participant demographics broken down by mobility aid, including age, gender, and travel frequency.}
    \label{fig:demographics}
\end{figure*}

\begin{figure*}
    \centering
    \includegraphics[width=1\linewidth]{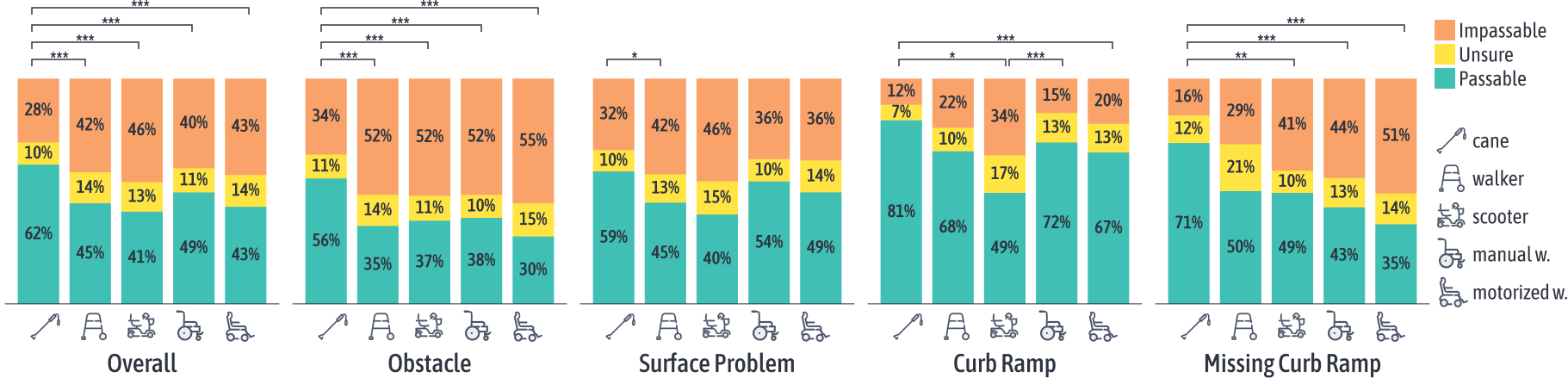}
    \caption{Passability assessment results for all 52 images and categories of \textit{obstacle}, \textit{surface problem}, \textit{curb ramp} and \textit{missing curb ramp}. Lines on top of bar charts represent results of analysis of variance based on mixed multinomial logistic regression, *$p<.05$, **$p< .01$, ***$p<.001$.}
    \Description{This is a data visualization of bar plots of passibility assessment results for 52 images all together and categories of obstacle, surface problem, curb ramp and missing curb ramp. Lines on top of bar charts represent results of analysis of variance based on mixed multinomial logistic regression.}
    \label{fig:image-selection-results}
\end{figure*}

\subsubsection{Survey Design}
Our survey had three parts: (1) study overview and background information, (2) image-based sidewalk passability rating and pair-wise comparisons, and (3) a ranking of sidewalk factors. 
The full survey is available as a PDF in supplementary material and online at \href{https://sidewalk-survey.github.io/}{https://sidewalk-survey.github.io/}.

\textbf{Part 1: Background information.}
The survey began with a study description and informed consent. 
The opening page stated the study goal was \sayit{to understand how people using different mobility devices perceive barriers in urban environments} and explained the \$10 USD remuneration, as well as that participants could save their responses and return to the survey later. 
The survey then collected basic demographic and mobility aid information. 
If respondents indicated that they used multiple mobility aids, we asked which \sayit{they use more frequently when going outside your home}. 
We then asked an open-form question about \sayit{What are the most difficult sidewalk barriers you encounter [using that mobility aid]?}. 

\textbf{Part 2.1: Image-based passability ratings.}
Part 2 included two sub-parts: (2.1) image-based passability ratings and (2.2) pair-wise comparisons. 
In Part 2.1, participants were shown images of sidewalk barriers and asked to judge their passability. 
Specifically, for each image, we asked, \sayit{When using your [mobility aid], do you feel confident passing this?} (\autoref{fig:survey-part2-instructions}); 
participants could select \sayit{Yes,} \sayit{No,} or \sayit{Unsure}. 
For each image, we added a salient red dot highlighting the target of interest and instructed participants to focus on the red dot when responding. 
To help guide participants and gather consistent responses, Part 2.1 began with an interactive tutorial showing an example image with the prompt: \sayit{Imagine yourself encountering these situations in real life. Would you be able to pass by the barrier?} (\autoref{fig:survey-part2-instructions}a).
The interactive tutorial then provided a definition of passability (\autoref{fig:survey-part2-instructions}b). 

After presenting instructions, the survey showed participants individual sidewalk images and, for each, voted on passability. 
Images were grouped into the nine distinct sets based on sidewalk feature or barrier type, including \textit{curb ramps}, \textit{surface problems}, and \textit{obstacles}---\autoref{tab:label-categories} and \autoref{fig:image-dataset-grid}. 
To mitigate ordering effects, we randomized both the sequence of the nine image sets as well as the order of images within each set. After a single image set was completed, the results were used to compute dynamic pairwise comparisons, and the participant entered Part 2.2.

\textbf{Part 2.2: Pairwise comparisons}
For Part 2.2, participants were shown the same images from Part 2.1 but asked to compare them with the question: \sayit{When using your [mobility aid], which do you feel more confident passing?}---see \autoref{fig:survey-part2b-pairwise}. 
Participants could select the \sayit{left image,} \sayit{right image,} or \sayit{the same}. 
These visual comparison studies are becoming increasingly common in urban science to evaluate perceptions of safety~\cite{salesses_collaborative_2013}, bikeability~\cite{evans-cowley_streetseen_2014}, beauty~\cite{goodspeed_research_2017}, and more. 
Comparing all 52 images to one another, however, would require 1,326 pairwise comparisons---an intractable number. 
Thus, we developed a different strategy informed by~\cite{harker_incomplete_1987}: first, images were compared only within their image set from Part 2.1. 
Second, images marked \sayit{Yes} were placed into one comparison set, while those marked \sayit{No} were placed in another; images marked \sayit{Unsure} were placed in both. 
This grouping strategy allowed for a more nuanced analysis of \sayit{Unsure} responses and reduced the total number of comparisons. 
In total, each participant could make between 6 and 15 comparisons depending on their Part 2.1 responses. 
Once they completed pairwise comparisons of a given image set, participants began another set, starting again in a new Part 2.1 set. 
This process was repeated until all nine image sets were completed.

\textbf{Part 3: Ranking of sidewalk barriers.} 
Finally, we asked participants to rank-order nine types of sidewalk barriers drawn from the literature~\cite{kasemsuppakorn_personalised_2009,kasemsuppakorn_understanding_2015,saha_project_2019, wheeler_personalized_2020} as well as from ADA guidelines~\cite{us_department_of_justice_2010_2010}, including \textit{uneven sidewalk panels}, \textit{narrow sidewalks}, \textit{missing curb ramps}, and \textit{sand/gravel surfaces} (\autoref{fig:survey-part3-ranking}). 
To aid comprehension, each barrier was accompanied by an example image (viewable by clicking on the image icon), and we randomized the initial rank-order options. 

\textbf{Survey completion.}
The survey ended with a thank-you and contact address. Participants who reported using multiple mobility aids in Part 1 were given the option to re-take Part 2 for their other selected aid(s). These participants could complete the additional survey immediately, defer, or decline.

\begin{figure*}
    \centering
    \includegraphics[width=1\linewidth]{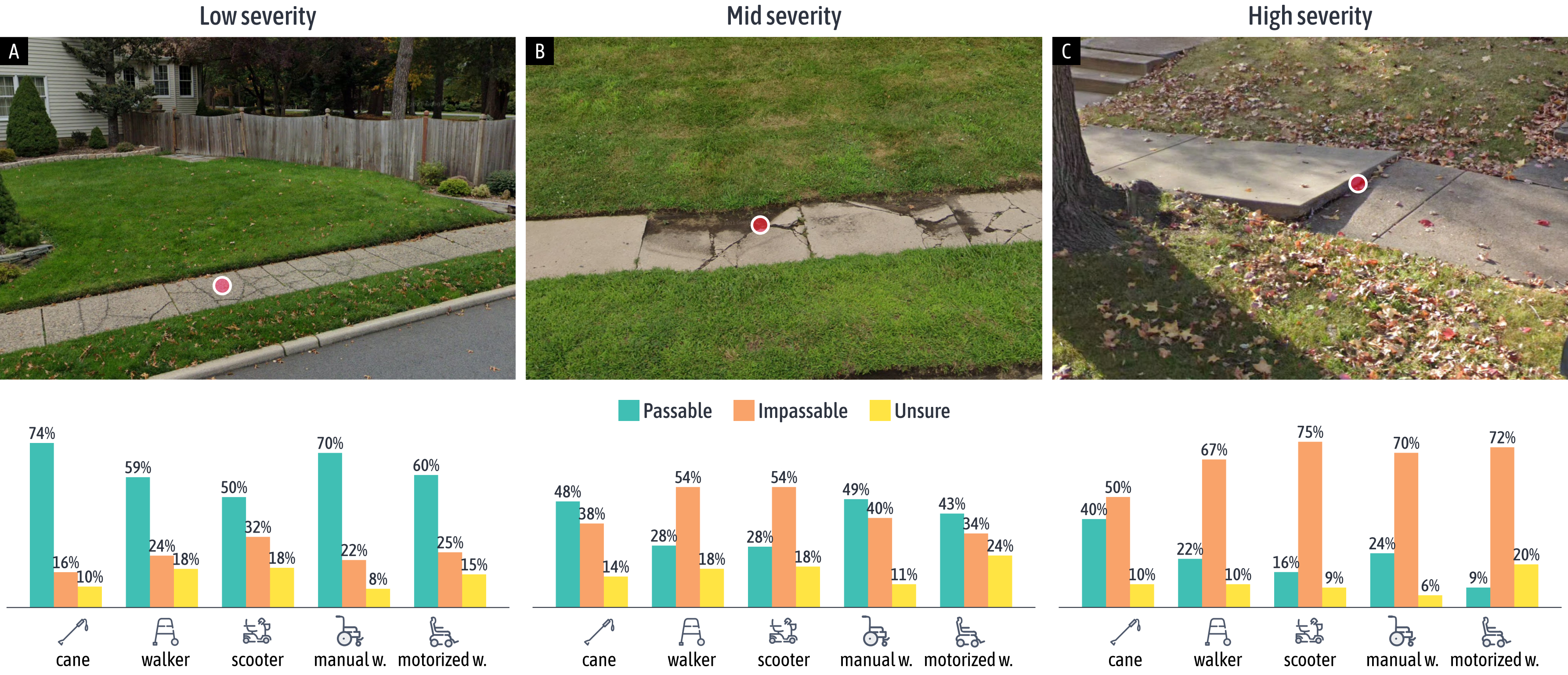}
    \caption{Passibility assessment results for category \textit{`cracks/broken surfaces + height differences'} on three severity levels of (A) low, (B) mid, and (C) high. While the assessment for both low and high severity are similar across mobility aids, we see a divide in the mid severity category, with walker and mobility scooters more likely to perceive them as impassible.}
    \label{fig:surface-problem-example}
    \Description{This figure is a data visualization of bar plots of passibility assessment results of category ‘cracks. Broken surfaces + height differences; on three severity levels of low, mid, and high. Images A,B,C show an example of each of the severity levels. While the assessment for both low severity and high severity are similar across mobility aids, we see a divide in the mid severity, with walker and mobility scooters more likely to perceive them as impassable.}
\end{figure*}

\subsection{Iterative Survey Development}
We designed the web survey in \textit{Figma} and implemented it in \textit{ReactJS v18.2}~\footnote{https://react.dev} (frontend) and \textit{Firebase v10.13.1}~\footnote{https://firebase.google.com/} (backend) hosted on \textit{GitHub Pages}. See our Github repo for details~\footnote{\href{https://github.com/makeabilitylab/accessibility-for-whom}{https://github.com/makeabilitylab/accessibility-for-whom}}. To design the survey, we used a human-centered, iterative process starting with with five rounds of internal testing amongst the research team and then four external pilots with mobility aid users. For the latter, we conducted a think-aloud session via Zoom and participants sharing their screens. Each session lasted for 60-90 minutes and participants were compensated \$30 for their time. 

Overall, pilot participants responded positively to the study topic, questions, and survey format. For example, pilot participant 3 (a manual wheelchair user) stated, \sayit{It’s very simple, this medium really helps because it’s also like a quick question, and quick decision[s] you have to make in real time. So you're triggering this similar emotional response that would be triggered from a real-life scenario.} Based on pilot study findings, we: 
(1) improved ranking question design with example images; 
(2) refined image sets to avoid overly similar or unclear images; 
(3) clarified terminology (\textit{e.g.,} "manholes" instead of "utility panels"); 
(4) implemented UI improvements, including increased font sizes for explainer texts and exit confirmation alerts.

\subsection{Survey Advertising and Recruitment}
Participants were recruited through disability organizations, social media, and word of mouth. 
Study advertisements linked to a screener, which asked about demographics, mobility aid use, and vision loss.
The screener questions and response options were designed based on prior surveys ~\cite{us_department_of_transporation_national_2022, the_university_of_kansas_national_2018, government_of_canada_canadian_2022, guensler_sidewalk_2017}.
See supplementary material for the complete list of screener questions.
We filtered for adults (18+) who use a mobility aid (\textit{walking cane}, \textit{walker}, \textit{mobility scooter}, \textit{manual wheelchair}, and \textit{motorized wheelchair}). 
Because our survey relied on a visual examination of images, we also excluded users of screen readers. 
The survey was posted for about two months in the summer of 2024. 

Similar to other recent online surveys~\cite{lawlor_suspicious_2021,griffin_ensuring_2022}, we experienced problems with fraudulent sign-ups. To mitigate this, we filtered screening responses based on IP address (no duplicates) as well as the respondent's qualitative descriptions of a prompt image (\textit{N=}5,239). 
In addition, we filtered out a smaller number of actual survey responses (\textit{N=}68) based on a combination of improbable completion times, IP and email address duplication, small screen sizes (making it difficult to see images), and whether the open-form responses seemed to be AI-generated (\textit{e.g.,} nonsensical responses).

\begin{figure*}
    \centering
    \includegraphics[width=1\linewidth]{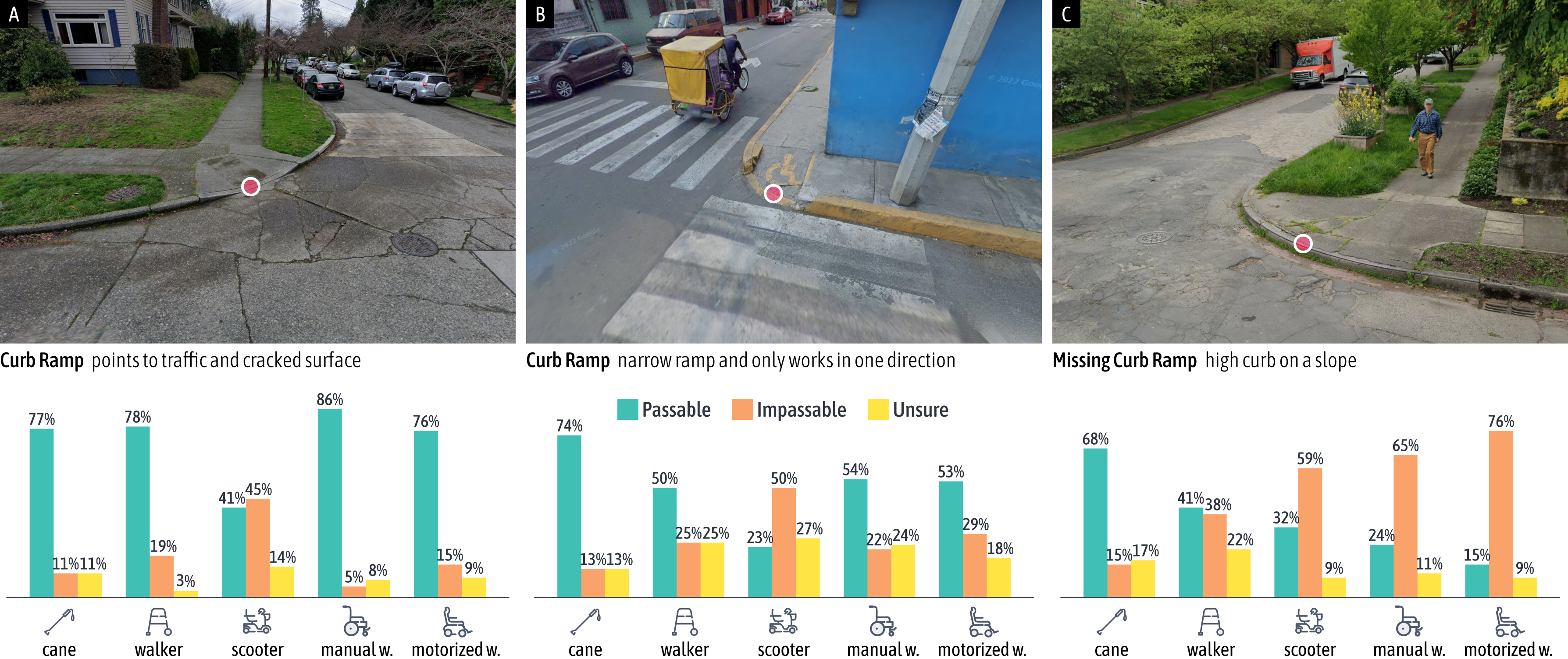}
    \caption{Passibility assessment results for category curb ramps (A \& B) and missing curb ramps (C). From the bar charts, we can see poorly designed or badly maintained curb ramps seem to be restrictive for people using mobility scooters. For missing curb ramps, wheeled mobility users perceive them as more challenging than walking cane and walker users.}
    \label{fig:curb-ramps-example}
    \Description{This figure is a data visualization of bar plots of passibility assessment results  for category curb ramps (image A \& B) and missing curb ramps (image C). From the bar charts we learned that poorly designed or badly maintained curb ramps come across as restrictive for people using mobility scooters. As for missing curb ramps, wheeled mobility users perceive them more challenging than walker and walking cane users}
\end{figure*}

\subsection{Data and Analysis}
We describe our analysis approach for each part of the survey. For the open-form questions, we employed a qualitative open coding method~\cite{charmaz_constructing_2006}, where one researcher developed a set of deductive themes based on the sidewalk barrier categories, then coded the responses accordingly. 
For survey Part 2.1, the passability assessment, we calculated the counts and the percentages of \say{Yes,} \say{No,} and \say{Unsure} votes for each mobility aid group. 
To investigate the differences in perceived passibility between different mobility groups, we conducted an analysis of variance based on mixed multinomial logistic regression, implemented using the multinomial-Poisson transformation~\cite{baker_multinomial-poisson_1994, guimaraes_understanding_2004, chen_note_2001}.  \textit{Post hoc} pairwise comparisons were conducted using the multinomial-Poisson transformation~\cite{baker_multinomial-poisson_1994} and corrected with Holm’s sequential Bonferroni procedure~\cite{holm_simple_1979}.

For Part 2.2, the pairwise comparison, we used Q score~\cite{salesses_collaborative_2013}, which is commonly employed in urban science literature for street-scene comparisons ~\cite{zhang_measuring_2018, goodspeed_research_2017, salesses_collaborative_2013, evans-cowley_streetseen_2014}. 
While some prior work uses a basic Win Ratio statistic~\cite{goodspeed_research_2017}, Q Score can accommodate tie scenarios, which, in our case, included pairwise results where the participant selected \say{Unsure.} 
Q score enhances Win Ratio of a certain image by incorporating the average Win Ratio of images it was preferred over, while subtracting the average loss ratios of images that were chosen over it~\cite{salesses_collaborative_2013}. 
Q score ranges from 0 to 10, with 10 indicating the most passable image and 0  the least (for that image set). 
To analyze our Q score data, we ranked the Q scores within each image subcategories and mobility group, and then conducted mixed-effects ordinal logistic regression on ranks~\cite{hedeker_random-effects_1994}.

Finally, we used Kemeny-Young rank aggregation~\cite{young_condorcets_1988, young_optimal_1995, kemeny_mathematics_1959} to analyze the rank order question in Part 3. Kemeny-Young is a prominent rank aggregation method in social choice theory~\cite{hamm_computing_2021}; it is based on the Kendall's Tau distance between rankings and outputs a consensus ranking that minimizes the sum of distances to the input rankings. 
\section{Findings}
\label{section:findings}

In total, we received 190 completed valid responses with a median completion time of 24.5 minutes.
An additional six participants began the survey but did not complete it (drop out rate of 3.2\%). Below, we begin by describing participant demographics before organizing our findings around the survey parts: passability, pairwise comparisons, and rank-ordering. 
We intermix quotes from the qualitative data to complement the quantitative findings. 

\subsection{Participant Demographics}
\label{section:}
A total of 144 individuals participated in the survey, with 34 participants providing responses for multiple mobility aids, resulting in 190 total responses. 
Participants were spread across age categories but leaned younger: 40\% were aged 18–34, 29\% were 35–54, 27\% were 55-74, and 4\% (\textit{N=}5) were 75-94. 
A slight majority identified as women 51\% (\textit{N=}73), 42\% as men, and 8\% as non-binary. Our participants tended to traveled frequently outside the home, most commonly 7+ times a week (38\%) followed by 4-6 times a week (36\%), 1-3 times a week (20\%), while 6\% (\textit{N=}8) traveled 1-3 times a month or less. 
Among the 190 responses, 53 reported using walking canes (28\%), 37 manual wheelchairs (20\%), 34 motorized wheelchairs (18\%), 32 walkers (17\%), 22 mobility scooters (12\%), and 12 others (6\%) such as crutches, rollators, and knee scooters. \autoref{fig:demographics} shows demographics by mobility aid.

\begin{figure*}
    \centering
    \includegraphics[width=1\linewidth]{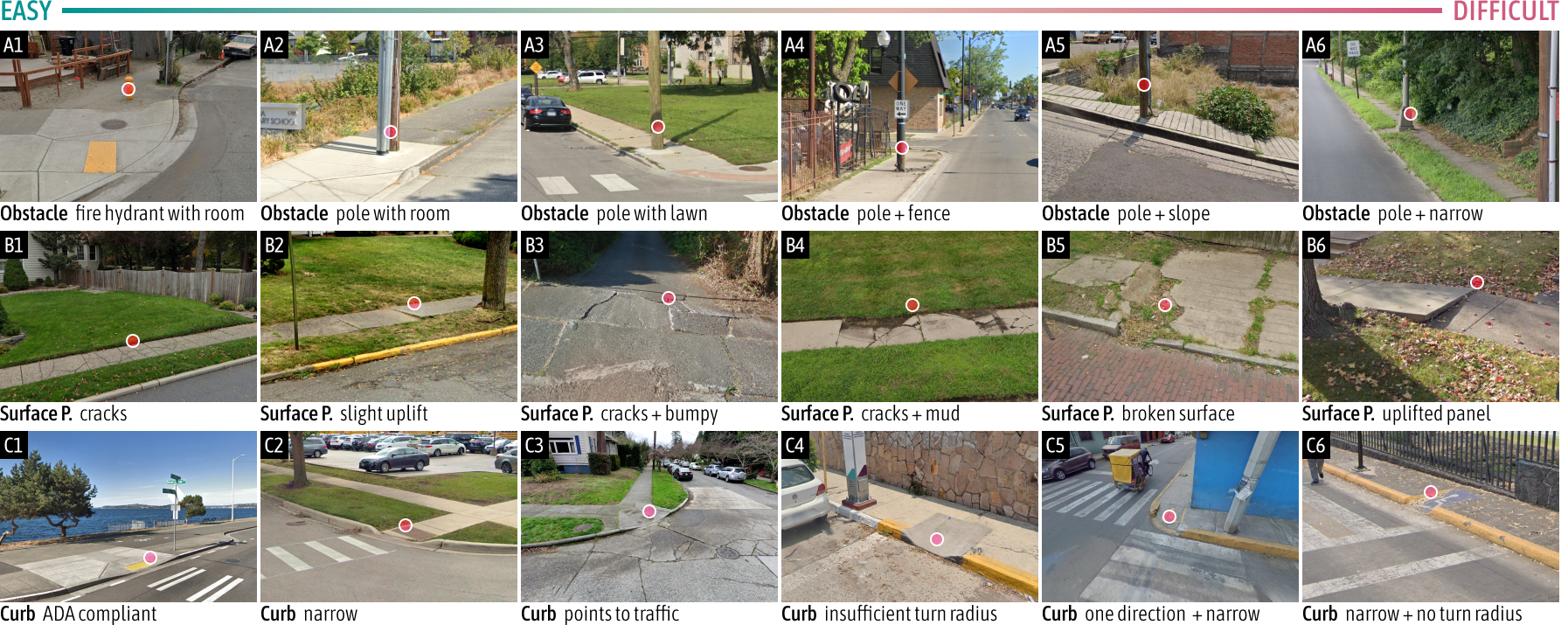}
    \caption{Image comparison result for subcategories of “fire hydrant and poles” (images A1-A6), “cracks and height difference” (images B1-B6), and “curb ramps” (images C1-C6). The images are ordered from left to right based on their Q score. For “fire hydrant and poles”, all mobility groups ranked in the order of A1 to A6 expect cane user group reversed the order of A5 and A6. For cracks and height difference", different groups varied slightly on their rankings for B2/B3 and B5/B6. For “curb ramps”, different groups varied slightly on their choices for C3/C4 and C5/C6. See the full Q score rankings for each group in the Appendix (\autoref{fig:pairwise-result}).}
    \label{fig:pairwise-result}
    \Description{This figure shows the image comparison result for subcategories of “fire hydrant and poles” (images A1-A6); “cracks and height difference” (images B1-B6); and “curb ramps” (images C1-C6). The images are ordered from left to right based on their Q score. For “fire hydrant and poles”, all mobility groups ranked in the order of A1 to A6 expect walking cane user group reversed the order of A5 and A6. For cracks and height difference", different groups varied slightly on their rankings for B2/B3 and B5/B6. For “curb ramps”, different groups varied slightly on their choices for C3/C4 and C5/C6.}
\end{figure*}

\subsection{Passability Assessment}
\label{section: passibility-assessment}
Our analysis of variance revealed significant main effects of mobility aid ($\chi^2(8, N=9,256) = 133.23, p < .001$) and barrier type ($\chi^2(6, N=9,256) = 16.08, p = .013$) on perceived passability. There was also a significant interaction between mobility aid and barrier type ($\chi^2(24, N=9,256) = 105.08, p < .001$).
Overall, mobility scooter users selected the highest number of impassable images---nearly half of all 52 sidewalk barriers shown were deemed impassable (46\%)---followed by users of motorized wheelchairs (43\%), walkers (42\%), manual wheelchairs (40\%), and walking cane users (28\%). Walking cane user responses differed significantly from all the other groups (\autoref{fig:image-selection-results}). Below, we describe our results as a function of high-level barrier type.

\textbf{Obstacles.}
Perceptions of sidewalk obstacle significantly differed among mobility aid groups. 
Walking cane users consistently rated obstacle images as more passable compared to users of walkers, mobility scooters, manual wheelchairs and motorized wheelchairs (all $p <$ .001, \autoref{fig:image-selection-results}). 
In our open-ended question about the most challenging sidewalk barriers, participants frequently cited obstacle-related issues, including parked cars/scooters/bikes, overgrown vegetation, signs, poles, traffic cones, construction, trash cans, \textit{etc}. 
One manual wheelchair user described a common challenge: \sayit{Stationary objects (typically a tree, planter, or pole) residing in the center of the walkway area... while people walking on foot can easily walk around these objects on either side, as a wheelchair user there is almost never enough room for me to pass on either side. This causes me to have to get assistance to get pushed either through a grassy lawn on one side (if this is even an option), or off the high curb and into the street to surpass the obstacle. All of these present major safety risks.}

\textbf{Surface problems.}
In an open-form question asking participants about the most challenging sidewalk barriers that they encountered, \textit{surface problems} were most frequently cited across all groups, contributing to 40\% (185/465) of total issues mentioned. 
Examples include unevenness, cracks, potholes, broken tiles, and damage caused by tree roots. 
A mobility scooter user stated: \sayit{Cracks, potholes, and uneven surfaces can cause instability or even damage to the scooter.}
Similarly, a manual wheelchair user shared the ongoing discomfort caused by surface problems:\sayit{the cracks/seams between sidewalk segments cause a constant, uncomfortable bumping that also shakes my legs off the wheelchair's footplate. Roads are kept smooth. Sidewalks are not.} 

In the image assessment portion of the survey, 7 of 22 surface problems (31\%) were selected as impassable across all groups. However, similar to obstacles, cane users found surface problems significantly more manageable than users of a walker ($p < .05$). 
When examining the three levels of severity within the subcategory of \textit{cracks/broken surfaces + height differences}, we observed that assessment for both low and high severity are similar across mobility aids. 
However, there was a divide in mid severity, with walker and mobility scooters more likely to perceive them as impassible and others as passable (\autoref{fig:surface-problem-example}).

\textbf{Curb ramps.}
Mobility scooters reported the most difficulty with \textit{using} low-quality curb ramps, finding ramps significantly more challenging compared not only to walking cane users ($p< .05$) but also to manual wheelchairs users ($p< .001$). For example, the diagonal curb ramp with a cracked surface in \autoref{fig:curb-ramps-example}A and the unidirectional, narrow curb ramp in \autoref{fig:curb-ramps-example}B received a low percentage of passable votes from mobility scooter users (23\% and 41\%, respectively), while other groups find these ramps to be more passable than not. However, our findings also highlight how \textit{all} mobility aid users are affected by poorly designed curb ramps. As one motorized wheelchair user stated: \sayit{When they [curb ramp] are too steep, the wheelchair gets stuck and the tires just spin.} A manual wheelchair user identified poorly maintained curb cuts as the most challenging barrier, noting, \sayit{The ones that have shifted over the years are especially difficult. I have perform a wheelie to get over them.}

\textbf{Missing curb ramps.}
Walking cane users find missing curb ramps to be less challenging compared to mobility scooter users, manual and motorized wheelchair users (all $p < .01$). 
Missing curb ramps are particularly challenging for wheeled mobility devices. A manual wheelchair user states: \sayit{My manual wheelchair is made out of mountain bike parts. It's meant to be used `off roading' as mountain bikes are typically used. That being said, I still would struggle if there weren't curb cutouts.} Despite being less prohibitive for walking cane users, 22 of 53 participants still cited "missing curbs" as a major barrier, noting they \sayit{require a lot of energy} and \sayit{can pose tripping hazards}. Similarly, 14 of 32 walker users mentioned "missing curbs" as the most difficult
sidewalk barrier, emphasizing that without curb ramps \sayit{large step downs are very difficult}.

\subsection{Pairwise Comparisons}
In the pairwise comparison (Part 2.2), participants were asked to compare images within each of the nine sub categories. Interestingly, our findings show minimal differences across mobility aid groups (no statistical difference observed). For example, \autoref{fig:pairwise-result} (A1-A6) shows the Q score ranking within the \textit{“fire hydrant and pole”} obstacle subcategory, with images arranged from the easiest to the most difficult to pass (left to right). All mobility aid groups produced the same rankings with one small exception: the walking cane user group switched the order of the \textit{pole on a slope} (\autoref{fig:pairwise-result} A5) and the \textit{pole with narrow passage} (\autoref{fig:pairwise-result} A6). We found a similar trend for the other eight barrier subcategories, for example, see \textit{"cracks, broken surfaces and height difference"} in \autoref{fig:pairwise-result} B1-B6 and \textit{"curb ramps"} in \autoref{fig:pairwise-result} C1-C6.

These results indicate that while people’s assessment of individual barriers can differ significantly by mobility aid usage---as we found in in \autoref{section: passibility-assessment}---their comparative judgments of the easiest and most difficult obstacles to navigate are remarkably similar. For instance, a fire hydrant in the middle of the sidewalk with some room to pass on either side (\autoref{fig:pairwise-result} A1) is consistently perceived as more passable by all groups than a pole in the middle of a narrow sidewalk (\autoref{fig:pairwise-result} A6). We present the complete Q score ranking for all nine image groups in the Appendix (\autoref{fig:image-grid}).

\subsection{Barrier Rankings}
\begin{figure}[t]
    \centering
    \includegraphics[width=1\linewidth]{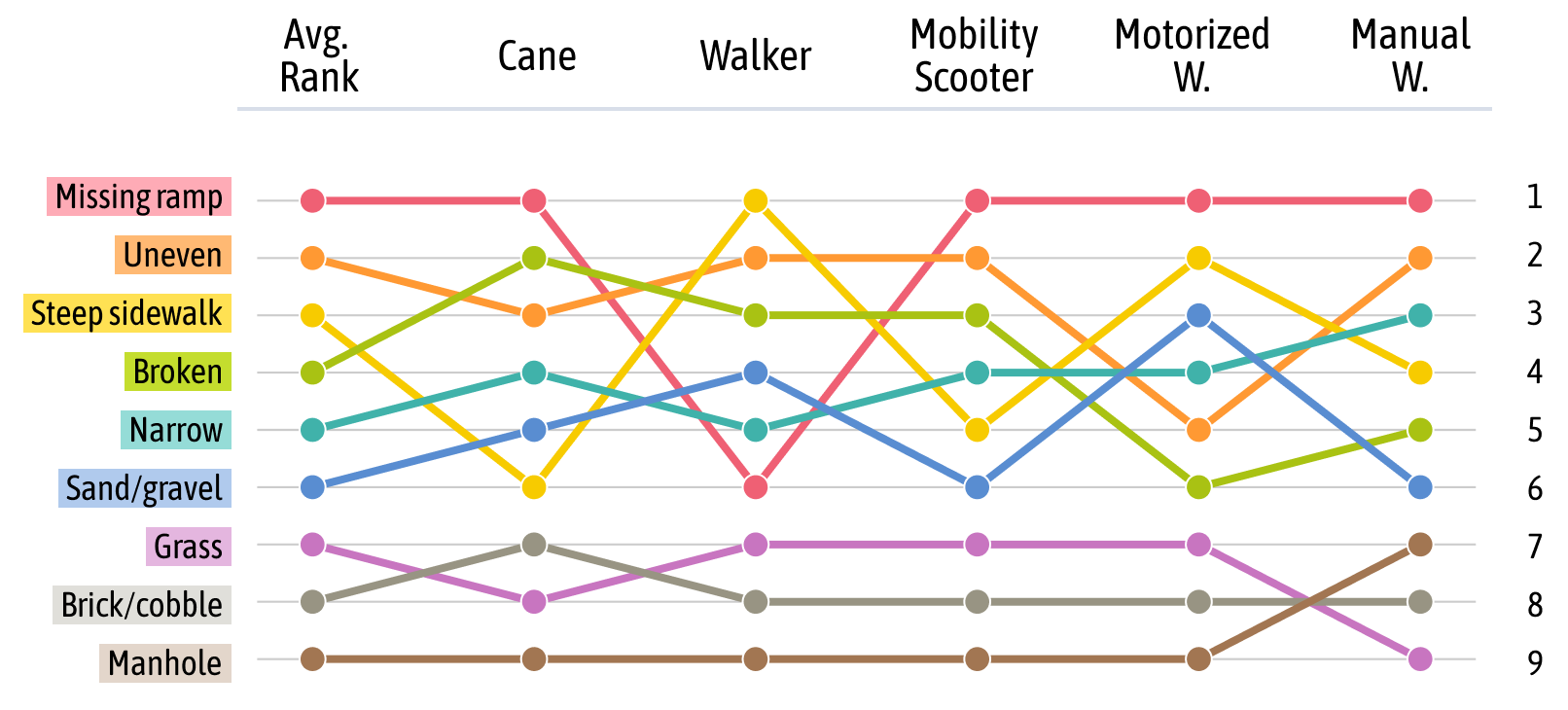}
    \caption{The rank-order results from Part 3 presented in a bump-chart. All groups except \textit{walking cane} rank missing curb ramps as most challenging. All five groups ranked grass surface, brick/cobblestone, and manhole covers (utility panels) as the least challenging barrier (though order changed slightly across groups).}
    \Description{This is an infographic image. It shows the rank order results from Survey Part 3 using the Kemeny-Young rank aggregation method. The barriers listed in the bump chart are sorted by the average rank position across groups from the most difficult (top) to the least (bottom). All groups except walking cane rank missing curb ramps as most challenging. All five groups ranked grass surface, brick/cobblestone, and manhole covers (utility panels) as the least challenging barrier (though the order changed slightly across groups).}
    \label{fig:survey-part3-ranking-results-chart}
\end{figure}

In the third and final section of the survey, participants ranked common sidewalk barriers from most difficult (Rank 1) to least (Rank 9)---see instructions in \autoref{fig:survey-part3-ranking}. 
\autoref{tab:survey-part3-ranking-results-table} shows the results as a table and \autoref{fig:survey-part3-ranking-results-chart} shows them as a bump chart; the barriers are sorted by average rank across the five mobility groups from most difficult (top) to least (bottom). Overall, the most difficult barriers were \textit{missing curb ramp} (ranked, on average, 2.4/9), \textit{uneven sidewalk panels} (3.2), and \textit{steep sidewalk slopes} (4.3), while \textit{grass surface} (7.4), \textit{brick/cobblestone} (7.7), and \textit{manhole/utility covers} (7.8) were perceived as least difficult. 

\begin{table}[t]
\centering
    \centering
    \renewcommand{\arraystretch}{1.2}
    \resizebox{\columnwidth}{!}{%
\begin{tabular}{@{}lrrrrrr@{}}
\textbf{Sidewalk barriers} &
  \textbf{\begin{tabular}[c]{@{}r@{}}Avg.\\ Rank\end{tabular}} &
  \textbf{Cane} &
  \textbf{Walker} &
  \textbf{\begin{tabular}[c]{@{}r@{}}Mobility\\ Scooter\end{tabular}} &
  \textbf{\begin{tabular}[c]{@{}r@{}}Manual\\ W.\end{tabular}} &
  \textbf{\begin{tabular}[c]{@{}r@{}}Motorized\\ W.\end{tabular}} \\ \midrule
Missing curb ramp &
  \cellcolor[HTML]{FAC3B2}2.4 &
  \cellcolor[HTML]{FDE1AF}4.4 &
  \cellcolor[HTML]{F9B7B3}1.6 &
  \cellcolor[HTML]{F9B8B3}1.7 &
  \cellcolor[HTML]{FAC4B2}2.5 &
  \cellcolor[HTML]{F9BAB3}1.8 \\
Uneven sidewalk panel &
  \cellcolor[HTML]{FBCFB1}3.2 &
  \cellcolor[HTML]{F9BEB3}2.1 &
  \cellcolor[HTML]{FAC1B2}2.3 &
  \cellcolor[HTML]{FAC9B2}2.8 &
  \cellcolor[HTML]{EAE991}6.6 &
  \cellcolor[HTML]{F9BDB3}2.0 \\
Steep slope &
  \cellcolor[HTML]{FDE0AF}4.3 &
  \cellcolor[HTML]{F9B7B3}1.6 &
  \cellcolor[HTML]{FEE7AE}4.8 &
  \cellcolor[HTML]{F8EFA4}5.8 &
  \cellcolor[HTML]{FBD3B1}3.5 &
  \cellcolor[HTML]{FAEFA6}5.7 \\
Broken surface/cracks &
  \cellcolor[HTML]{FDE1AF}4.4 &
  \cellcolor[HTML]{FBCDB1}3.1 &
  \cellcolor[HTML]{F9BAB3}1.8 &
  \cellcolor[HTML]{FCDAB0}3.9 &
  \cellcolor[HTML]{E1E685}7.1 &
  \cellcolor[HTML]{F3ED9D}6.1 \\
Narrow sidewalk &
  \cellcolor[HTML]{FDE4AF}4.6 &
  \cellcolor[HTML]{FDE0AF}4.3 &
  \cellcolor[HTML]{FBCCB1}3.0 &
  \cellcolor[HTML]{FAEFA6}5.7 &
  \cellcolor[HTML]{FCF0A9}5.6 &
  \cellcolor[HTML]{FDE6AF}4.7 \\
Sand/gravel &
  \cellcolor[HTML]{FEEDAE}5.2 &
  \cellcolor[HTML]{FCD8B0}3.8 &
  \cellcolor[HTML]{FDE6AF}4.7 &
  \cellcolor[HTML]{EAE991}6.6 &
  \cellcolor[HTML]{FCDDB0}4.1 &
  \cellcolor[HTML]{E8E88E}6.7 \\
Grass surface &
  \cellcolor[HTML]{DCE47E}7.4 &
  \cellcolor[HTML]{F5ED9F}6.0 &
  \cellcolor[HTML]{DFE582}7.2 &
  \cellcolor[HTML]{D5E174}7.8 &
  \cellcolor[HTML]{E1E685}7.1 &
  \cellcolor[HTML]{C3DA5C}8.8 \\
Brick/cobblestone &
  \cellcolor[HTML]{D7E277}7.7 &
  \cellcolor[HTML]{E7E88C}6.8 &
  \cellcolor[HTML]{E7E88C}6.8 &
  \cellcolor[HTML]{BFD857}9.0 &
  \cellcolor[HTML]{D1DF6F}8.0 &
  \cellcolor[HTML]{D3E072}7.9 \\
Manhole covers &
  \cellcolor[HTML]{D5E174}7.8 &
  \cellcolor[HTML]{DFE582}7.2 &
  \cellcolor[HTML]{DAE37B}7.5 &
  \cellcolor[HTML]{CADD66}8.4 &
  \cellcolor[HTML]{C7DB61}8.6 &
  \cellcolor[HTML]{E1E685}7.1 \\ 
\end{tabular}%
}
    \caption{The rank-order results from Part 3 using the Kemeny-Young rank aggregation method. The barriers are sorted by the average rank position across groups from the most difficult (top) to the least (bottom). Table cells are colored in pink (most difficult) and green (least difficult) scale.}
    \label{tab:survey-part3-ranking-results-table}
\end{table}

\textit{Missing curb ramps} were consistently ranked as the most challenging barrier except by walking cane users, who ranked it sixth and, instead, ranked \textit{steeply sloped sidewalks} first. As one manual wheelchair user said: \sayit{high curbs without ramps} is the most difficult barrier as it \sayit{forces me to seek for alternative routes}. 
Interestingly, all five groups rated \textit{grassy sidewalks}, \textit{brick/cobblestone surfaces}, and \textit{manhole/utility panel covers} as least difficult, although there was a slight variation in the exact order across groups. 
While \textit{steep sidewalks} posed significant problems to cane users (Rank 1) and manual wheelchairs (Rank 2), they were less challenging to the powered aids: motorized wheelchairs (Rank 4) and mobility scooters (Rank 5), as well as the walkers (Rank 6). 
It is not just the uphills that are impediments---requiring significant strength and endurance to overcome, but also the downhill slopes. 
As one rollator user commented: \sayit{the sharp downhill declines require a lot of braking and care.} The mobility scooter rankings and motorized wheelchair rankings are most similar but differ slightly in their ranking of \textit{path narrowness} (Rank 4 \textit{vs.} 3), \textit{broken surfaces} (Rank 3 \textit{vs.} 5), and \textit{sidewalk steepness} (Rank 5 \textit{vs.} 4).

\section{Applications}
\label{section:applications}
We now demonstrate how our survey findings can be used to create accessibility-oriented analytical maps and personalized routing algorithms. We first synthesize our  findings into user preferences before describing our two prototypes.

\subsection{User Preferences}
While ~\autoref{section:findings} was largely organized around barrier types, here we summarize findings by mobility aid. 
Our intent is to provide a more holistic synthesis across different survey parts and demonstrate how this data can be used to create more personalized, disability-infused mapping applications.

\begin{figure}[b]
    \centering
    \includegraphics[width=1\linewidth]{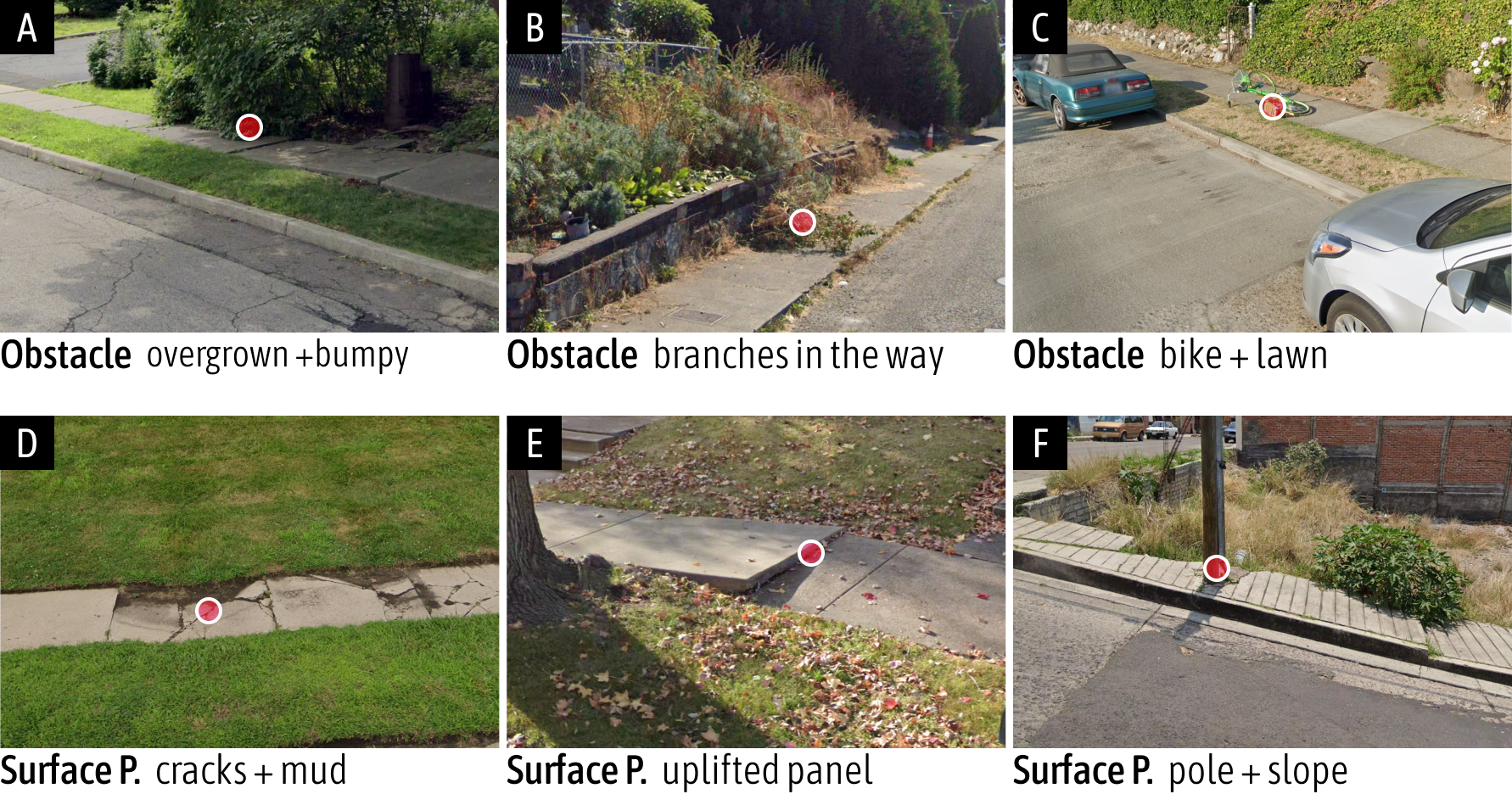}
    \caption{Examples of the least passable images across mobility groups.}
    \Description{This figure shows an array of six examples of the least passable images for each mobility group.}
    \label{fig:least-passable}
\end{figure}

\textbf{Walking canes.}
Walking cane users generally showed more confidence in maneuvering through or around sidewalk barriers compared to other groups. However, they still perceive high severity obstacles and high severity surface problems to be challenging (37\% and 44\% passable votes, respectively). 
The top two most difficult sidewalk barriers for walking cane users were overgrown vegetation on an already narrow sidewalk and branches obstructing the walkway (\autoref{fig:least-passable}A and B), with only 19\% and 23\% of users, respectively, indicating they could confidently pass.

\begin{figure*}
  \centering
  \includegraphics[width=\linewidth]{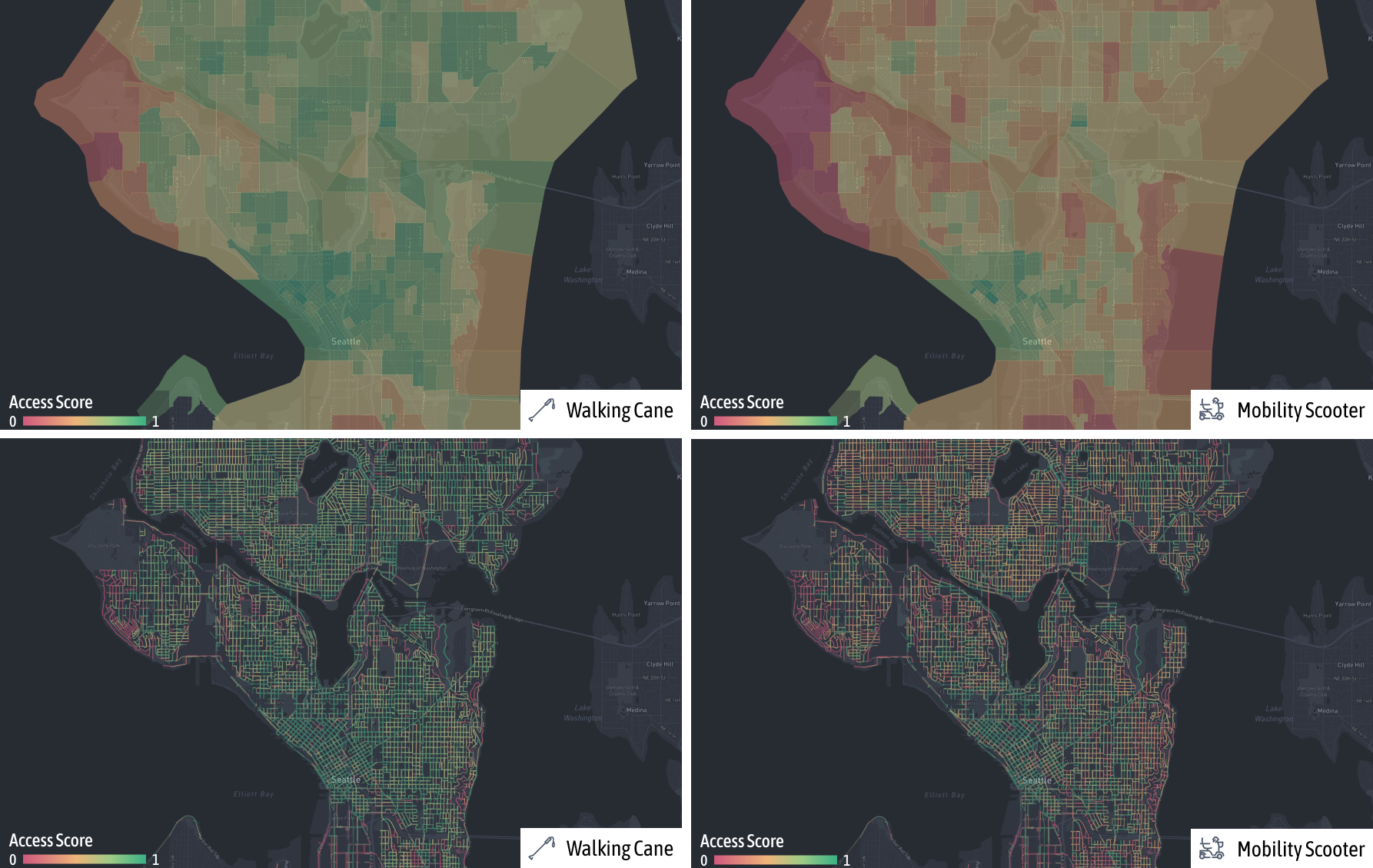}
  \caption{AccessScore maps visualizing sidewalk quality in Seattle for two groups: walking cane and mobility scooter (red is least accessible; green is most). Top two shows AccessScore by neighborhood; bottom two shows AccessScore by sidewalk segment. From the comparisons between walking cane users and mobility scooter users, we can see while downtown area may be equally accessible for both user groups, other areas are less accessible for mobility scooter users. }
  \Description{This figure shows AccessScore maps visualizing sidewalk quality in Seattle. Top two shows AccessScore by neighborhood; bottom two shows AccessScore by sidewalk segment. From the comparisons between walking cane users and mobility scooter users, we can see while downtown area may be equally accessible for both user groups, other areas are less accessible for mobility scooter users.}
  \label{fig:fig:access-maps}
\end{figure*}

\textbf{Walkers.}
Walker users were particularly sensitive to narrow sidewalks, including sidewalks narrowed by obstacles such as vegetation (40\% of walker users voted passable), parked cars, scooters, and bikes (32\%), as well as inherently narrow sidewalk surfaces (32\%). 
People who use walkers also struggle with cracks and uneven sidewalks, with more than 45\% of the votes indicating they are difficult to pass. 
The most challenging barriers for walker users were a parked bike in the middle of the sidewalk and branches obstructing the walkway (\autoref{fig:least-passable}C and B), with only 9\% and 10\% of users, respectively, indicating they could pass these obstacles.

\textbf{Mobility scooters.}
Mobility scooter users marked the most images as impassable (24 of 52 images). 
Examining users' passability confidence across severity levels revealed that these users were more likely to find images in both mid- and high-severity levels impassable, with only a 55\% passable ratio. 
This is lower compared to all other mobility aid users: walking cane (74\% ), walker (58\%), manual wheelchair (68\%), and motorized wheelchair (59\%).
Mobility scooter users were also particularly sensitive to poorly designed curb ramps, with a low passibility rate for curb ramps of 49\%. 
The top three most difficult sidewalk barriers for mobility scooter users were overgrown vegetation on a narrow sidewalk (\autoref{fig:least-passable}A), a broken sidewalk surface with mud (\autoref{fig:least-passable}D), and an uplifted sidewalk panel due to tree roots (\autoref{fig:least-passable}E), each with only 14\% of users indicating they could pass these barriers.

\textbf{Manual wheelchairs.}
Manual wheelchair users found high severity obstacles (18\% passable), surface problems (29\% passable), and all missing curb ramps (24\% passable) to be particularly challenging. 
Their top two most difficult sidewalk barriers were overgrown vegetation on a narrow sidewalk (\autoref{fig:least-passable}A) and a pole in the middle of the sidewalk with slope (\autoref{fig:least-passable}F), with only 11\% of users indicating they could pass these obstacles for both barriers.

\textbf{Motorized wheelchairs.}
Motorized wheelchair users showed similar patterns to manual wheelchair users but were even more sensitive to missing curb ramps (20\%  passable). This echoed an insight from one of our pilot participants: \sayit{If I am on a manual wheelchair and I see a missing curb ramp, I can do a wheelie to get on top of it, but it might not be possible when using a motorized wheelchair.} The top two most difficult sidewalk barriers for motorized wheelchair users were overgrown vegetation on a narrow sidewalk (\autoref{fig:least-passable}A) and a parked bike in the middle of the sidewalk (\autoref{fig:least-passable}C), with only 6\% of users indicating they could pass these obstacles for each barrier.

\begin{figure*}
  \centering
  \includegraphics[width=\linewidth]{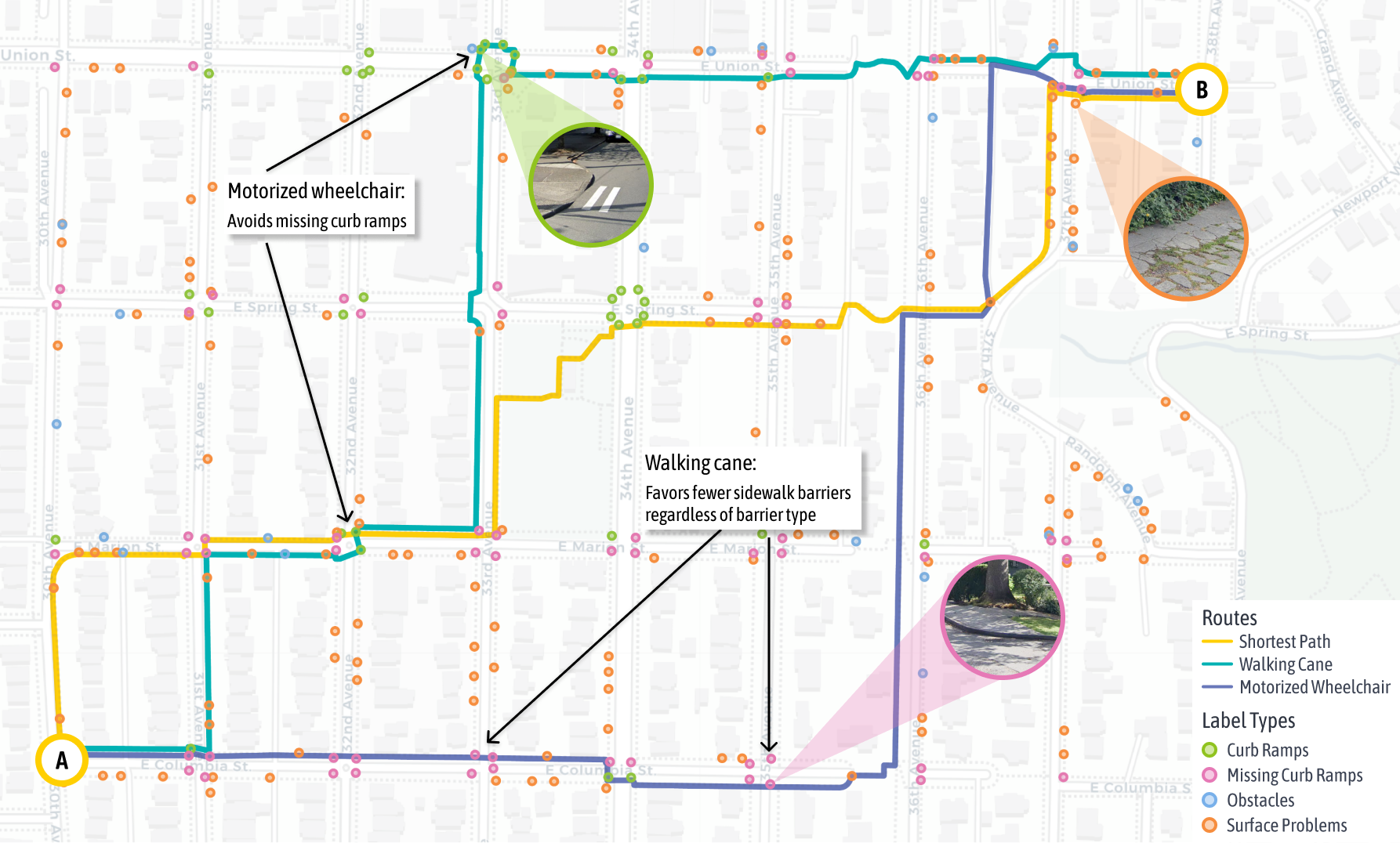}
  \caption{Routing application using OSMnx to generate routes between A \& B based on our survey data. Yellow route shows the absolute shortest path; teal shows the route for walking cane, this route favours fewer sidewalk barriers regardless of category; purple shows the route for motorized wheelchair, this route avoids missing curb ramps at all costs. When hovering over the labels, users can see what the sidewalk issues look like in streetview.}
  \Description{This figure shows routing application using OSMnx to generate routes between A \& B based on user preferences. Yellow route shows the absolute shortest path; teal shows the route for walking cane, this route favors fewer sidewalk barriers regardless of category; purple shows the route for motorized wheelchair, this route avoids missing curb ramps at all costs.}
  \label{fig:routing}
\end{figure*}

\subsection{Accessibility Map}

High-quality sidewalks play a vital role in the urban environment by encouraging physical activity~\cite{lopez_obesity_2006}, facilitating connectivity~\cite{randall_evaluating_2001}, increasing safety~\cite{abou-senna_investigating_2022}, and enhancing the sense of community~\cite{demerath_social_2003, bise_sidewalks_2018}. 
Current commercial tools like Walk Score~\cite{walk_score_walk_2007} take into account the use of sidewalks in gaining access to important amenities, and have been widely used by people to make informed decisions about where to live and which transportation modes to use. 
However, these tools often fail to capture the nuances of sidewalk accessibility for people with varying levels of mobility. 
The same sidewalk infrastructure can present drastically different levels of quality and usability for mobility aid users.

To address this problem, we prototyped an urban analytic tool that showcases sidewalk quality based on different mobility aid groups using data from our survey.
We used Project Sidewalk open label dataset (curb ramps, missing curb ramps, obstacles, and surface problems) from Seattle\footnote{\href{https://seattle.projectsidewalk.org/api}{https://seattle.projectsidewalk.org/api}} and mapped the labels onto sidewalk geometry gathered from the \textit{Seattle Open Data Portal}\footnote{\href{https://data-seattlecitygis.opendata.arcgis.com/datasets/SeattleCityGIS::sidewalks-1/about}{https://data-seattlecitygis.opendata.arcgis.com/datasets/SeattleCityGIS::sidewalks-1/about}}. 
We extended previous methods of using Project Sidewalk~\cite{li_interactively_2018, hara_scalable_2014, li_pilot_2022} labels to calculate \textit{AccessScore} by incorporating our survey findings.
The confidence that a sidewalk barrier type is not passable ($C_{label}$) was determined using the percentage of \sayit{No} and \sayit{Unsure} votes from ~\autoref{fig:image-selection-results}. For example, $C_{SurfaceProblem}$ for walking cane users is $0.54$, thus we weighted surface problems by $0.54$ when calculating their \textit{AccessScore}.
We generated sidewalk accessibility maps at both segment and neighborhood scales, with scores ranging from 0 (least accessible) to 1 (most accessible). 

\autoref{fig:fig:access-maps} compares the results for walking cane and mobility scooter users. 
The results show that, while downtown Seattle may be accessible for both groups, mobility scooter users face more challenges in other geographic areas. 
Such visualizations act like a Walk Score~\cite{walk_score_walk_2007} for mobility aid users, they can \textit{help people in choosing suitable living locations} and \textit{guide officials in prioritizing accessibility improvements}.

\subsection{Personalized Routing}
Existing navigation tools (\textit{e.g.,} Google Maps, Apple Maps) fail to address the needs of people with mobility disabilities. This section demonstrates how "one-size-fits-all" applications are insufficient for people with different mobility aids and how our survey data enables more accurate personalized routing. 

To develop a routing prototype, we first created a topologically connected routable network for our study area using the sidewalk network from OSM.
We then integrated Project Sidewalk labels by mapping obstacles and surface problems onto sidewalk segments, and (missing curb ramps) were mapped onto the crossing segments. 
To incorporate user profiles, we again used the confidence score that a sidewalk barrier type is not passable ($C_{label}$). 
Then, for each segment in the sidewalk network, we calculated the weighted distance for each segment as the segment length plus $C_{label}$ multiplied by the number of labels and 10\% of the segment length~\cite{tannert_disabled_2018}. 
Using OSMnx~\cite{boeing_osmnx_2017}, we next calculated the shortest distance between two intersection points (30th Avenue and East Columbia Street; 38th Avenue and East Union Street in Seattle) based on these weighted distances.

~\autoref{fig:routing} shows the shortest paths using absolute length and weighted length for walking cane users and motorized wheelchair users, with Project Sidewalk labels overlaid on the map.
The results demonstrate that users are given different optimal paths based on their specific needs and preferences. 
Walking cane users are routed along a path with some missing curb ramps but almost free of surface problems and obstacles, while motorized wheelchair users are given a longer path that avoids all areas with missing curb ramps. 
The results powerfully demonstrate how leveraging crowdsourced accessibility data and user preferences can yield more \textit{accurate and personalized routing algorithms} for mobility aid users.

\section{Discussion}
Through our application of personalized accessibility maps and routing applications, we showed how data and insights from our survey findings can help inform the development of more accurate navigation and analytical tools. 
We now situate our findings in related work, highlight how this survey contributes to personalized routing and accessibility mapping for mobility disability groups as well as present directions for future research.

\subsection{Online Image Survey Method}
In this study, we conducted a large-scale image survey (\textit{N=}190) to gather perceptions of sidewalk barriers from different mobility aid user groups. 
This approach helped us to collect insights on the differences between mobility aid user groups as well as shared challenges.
Previous research exploring the relationship between mobility aids and physical environment have mainly employed methods including in-person interviews~\cite{rosenberg_outdoor_2013}, GPS tracking~\cite{prescott_exploration_2021, prescott_factors_2020,rosenberg_outdoor_2013}, and online questionnaires~\cite{carlson_wheelchair_2002}. While interviews and tracking studies typically yield rich detailed information, they are limited to a small sample size. Online text based questionnaires often achieve larger sample sizes but at a cost of depth and nuance. Our image survey method struck a balance between sample size and detail. We collected a large sample within a relatively short time frame, enabling us to gather valuable insights and synthesize patterns across user groups.

Despite advantages, our approach has some limitations. Although street view images help situate and ground a participant's response---as one pilot participant said ``\textit{You're triggering a similar response to a real-life scenario''}, they cannot fully replicate the experience of evaluating a sidewalk \textit{in situ}. The lack of physical interaction with the environment limits the assessment of certain factors. For instance, one of our pilot participants noted that determining whether they could navigate past an obstacle like a trash can varies depending on \sayit{whether the trash can is light enough so I can push it away.} Using our findings as a backdrop, future work should conduct follow-up interviews and in-person evaluations. Such approaches would complement the quantitative data with richer qualitative insights, allowing researchers to better understand the patterns observed in quantitative data as well as the reasoning behind mobility aids users’ assessment.

\subsection{Personalized Accessibility Maps}
Our approach to infuse accessibility maps and routing algorithms with personalized information contributes to the field of accessible urban navigation and analytics. 
Based on our findings, we implemented two accessibility-oriented mapping prototypes, which demonstrate how our data can be used in urban accessibility analytics and personalized routing algorithms. While our current implementation serves as a proof of concept, future research could explore using our findings with more advanced modeling methods such as fuzzy logic~\cite{kasemsuppakorn_personalised_2009, gharebaghi_user-specific_2021, hashemi_collaborative_2017} and AHP~\cite{kasemsuppakorn_personalised_2009,kasemsuppakorn_understanding_2015, hashemi_collaborative_2017}. 

For our current map applications, we used a single set of open-source sidewalk data from Project Sidewalk. However, we acknowledge that other important factors are not included, such as sidewalk topography, width, stairs, crossing conditions, paving material, lighting conditions, weather, and pedestrian traffic~\cite{rosenberg_outdoor_2013,kasemsuppakorn_personalised_2009,darko_adaptive_2022,hashemi_collaborative_2017,sobek_u-access_2006,bigonnesse_role_2018}. 
Future work should build upon our foundation by incorporating more crowdsourced and government official datasets.

While mobility aids play a crucial role in determining accessibility needs, we must recognize that individuals using the same type of mobility aid may have diverse preferences. As one of our pilot participants stated, \sayit{your wheelchair has to be shaped and fitted to your body similar to how you need shoes specifically for your feet.} This insight underscores the need for personalization beyond broad mobility aid categories. Other factors including age~\cite{rosenberg_outdoor_2013}, disability type~\cite{prescott_factors_2020}, body strength~\cite{prescott_factors_2020}, and route familiarity~\cite{kasemsuppakorn_understanding_2015} should be explored in the future. Our attempt in creating personalized maps is not to provide a one-size-fits-all solution for generalized mobility aid groups, but rather to leverage the power of defaults~\cite{nielsen_power_2005} and offer users an improved baseline from which they can easily customize based on their individual needs.

\subsection{Limitations and Future Work}
Due to the visual nature of our survey—images were the primary stimuli—we specifically excluded people who are blind or have low vision\footnote{That said, the custom online survey was made fully screen reader accessible; see \href{https://sidewalk-survey.github.io/}{https://sidewalk-survey.github.io/} for the images and alt text.}. However, as noted previously, many different disabilities can impact mobility, including sensory, physical, and cognitive. Prior research has explored the incorporation of visually impaired or blind individuals into route generation~\cite{volkel_routecheckr_2008}, recognizing shared barriers and the prevalence of multiple disabilities among users. Building upon this foundation, future work should expand the participant pool to include a broader range of disabilities, thereby providing a more comprehensive understanding of diverse accessibility needs.

While we demonstrated two basic scenario applications, our survey findings and personalized mapping approach have potential for broader implementation. One promising direction is in developing barrier removal strategies for policymakers~\cite{eisenberg_barrier-removal_2022}. Current government plans often rely on simple metrics, such as population density or proximity to public buildings~\cite{seattle_department_of_transportation_seattle_2021}. Our methodology could enhance these efforts by identifying sidewalk barriers whose removal would yield the greatest overall benefit to the largest percentage of mobility aid users in the form of connected, safe, accessible routes.
\section{Conclusion}
In conclusion, this study advances the understanding of how different mobility aid users perceive sidewalk barriers. By conducting an online image survey with five mobility aid groups, we gathered nuanced perspectives on how users of walking canes, walkers, mobility scooters, manual wheelchairs, and motorized wheelchairs navigate the urban environment. From the findings synthesized in this study, we created two example applications, \textit{i.e.,} an interactive accessibility rating maps based on each user group's perceived passability and a disability-aware routing prototype based on OSMnx to generate personalized, optimal paths for each mobility group. 
These applications showcase the potential for our survey data to inform people with mobility disabilities about residential and social choices, provide personalized route planning strategies, and develop analytical tools to identify obstacles and assess the impact of their removal for different user groups. 
\begin{acks}
We thank all participants who took part in this study, without whom this project would not have been possible. We also thank Jacob O. Wobbrock for his help with data analysis, Michael Saugstad, Zhihan Zhang, Minchu Kulkarni, Jerry Cao for their help in survey/visualization development, as well as academic writing advisor Sandy Kaplan, UW CREATE community enagagement manager Kathleen Quin Voss, and the Allen School Computer Science Laboratory Group. This work was supported by NSF SCC-IRG \#2125087.
\end{acks}

\bibliographystyle{ACM-Reference-Format}
\bibliography{references}

\newpage
\appendix
\section{Appendix}
\begin{figure}[ht]
    \centering
    \onecolumn\includegraphics[width=0.9\textwidth]{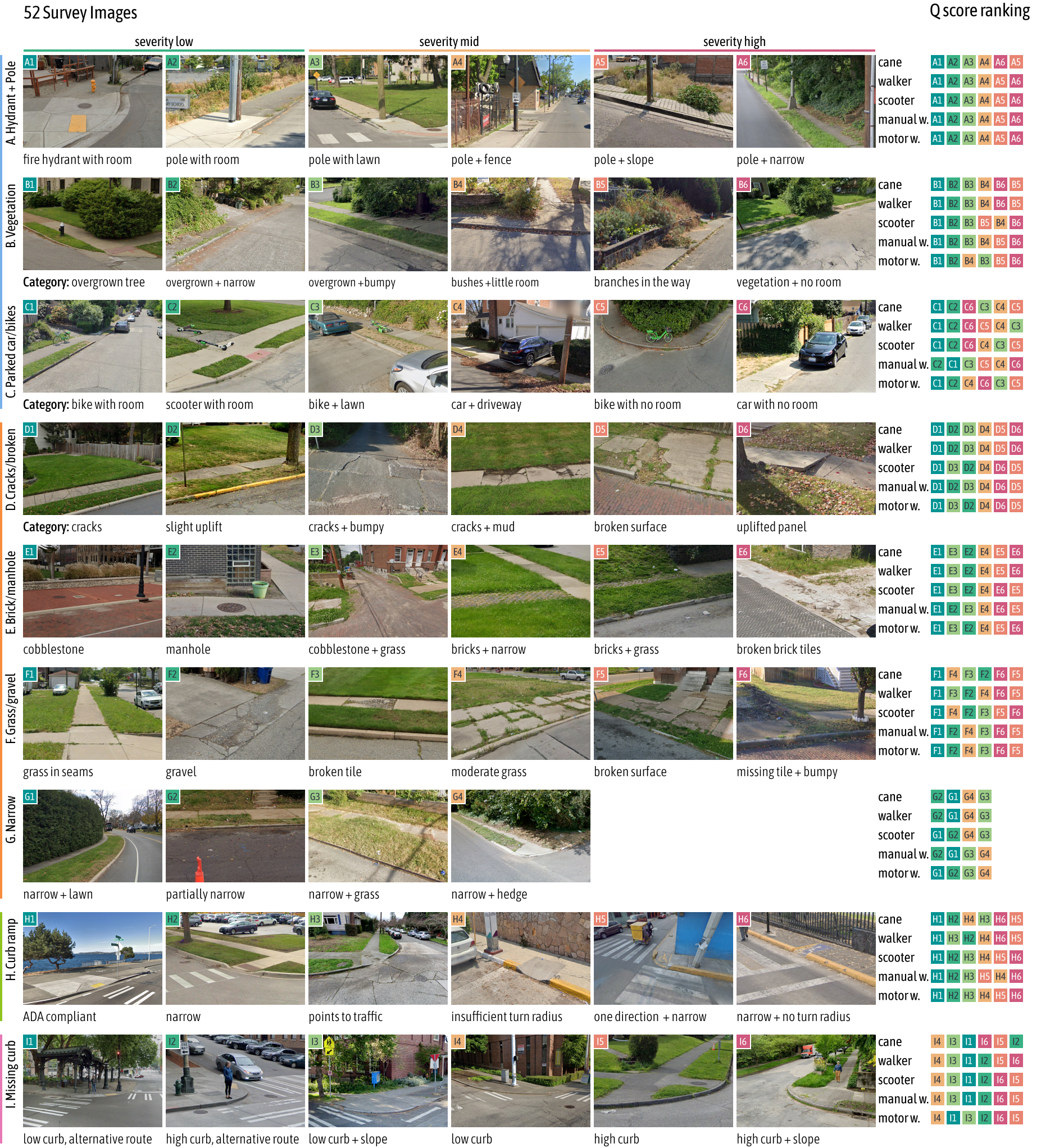}
    \caption{This figure presents 52 images used in the survey categorized into 9 sidewalk issue types (A-I). Each category contains 6 images, with 2 images per severity level (low, medium, high). The categories are further grouped into 4 Project Sidewalk label categories: Obstacles (A-C), Surface problems (D-G), Curb ramps (H), and Missing curb ramps (I).
    The matrix on the right displays the Q score ranking per user group for each category, arranged from the most passable (left) to the least passable (right).
    }
    \Description{This figure presents 52 images used in the survey categorized into 9 sidewalk issue types (A-I). Each category contains 6 images, with 2 images per severity level (low, medium, high). The categories are further grouped into 4 Project Sidewalk label categories: Obstacles (A-C), Surface problems (D-G), Curb ramps (H), and Missing curb ramps (I).
    The matrix on the right displays the Q-score ranking per user group for each category, arranged from the most passable (left) to the least passable (right).}
    \label{fig:image-grid}
\end{figure}

\begin{table*}[t]
\centering
\small 
\resizebox{\linewidth}{!}{%
\begin{tabular}{@{}p{6cm} c c c c p{4cm}@{}}
\toprule
\rowcolor[HTML]{FFFFFF} 
\textbf{Mobility Devices} &
  \multicolumn{1}{c}{\cellcolor[HTML]{FFFFFF}\textbf{Our Survey}} &
  \multicolumn{1}{c}{\cellcolor[HTML]{FFFFFF}\textbf{NHTS}} &
  \textbf{NSHD} &
  \textbf{CSD} &
  \textbf{Reasons For Exclusion} \\ \midrule
\rowcolor[HTML]{F3F3F3} 
Walking cane &
  \multicolumn{1}{c}{\cellcolor[HTML]{F3F3F3}\ding{51}} &
  \multicolumn{1}{c}{\cellcolor[HTML]{F3F3F3}\ding{51}} &
  \ding{51} &
  \ding{51} &
  \cellcolor[HTML]{FFFFFF} \\
\rowcolor[HTML]{FFFFFF} 
Walker &
  \multicolumn{1}{c}{\cellcolor[HTML]{FFFFFF}\ding{51}} &
  \multicolumn{1}{c}{\cellcolor[HTML]{FFFFFF}\ding{51}} &
  \ding{51} &
  \ding{51} &
  \cellcolor[HTML]{FFFFFF} \\
\rowcolor[HTML]{F3F3F3} 
Motorized scooter &
  \multicolumn{1}{c}{\cellcolor[HTML]{F3F3F3}\ding{51}} &
  \multicolumn{1}{c}{\cellcolor[HTML]{F3F3F3}\ding{51}} &
  \ding{51} &
  \ding{51} &
  \cellcolor[HTML]{FFFFFF} \\
\rowcolor[HTML]{FFFFFF} 
Manual wheelchair &
  \multicolumn{1}{c}{\cellcolor[HTML]{FFFFFF}\ding{51}} &
  \multicolumn{1}{c}{\cellcolor[HTML]{FFFFFF}\ding{51}} &
  \ding{51} &
  \ding{51} &
  \cellcolor[HTML]{FFFFFF} \\
\rowcolor[HTML]{F3F3F3} 
Motorized wheelchair &
  \multicolumn{1}{c}{\cellcolor[HTML]{F3F3F3}\ding{51}} &
  \multicolumn{1}{c}{\cellcolor[HTML]{F3F3F3}\ding{51}} &
  \ding{51} &
  \ding{51} &
  \multirow{-5}{*}{\cellcolor[HTML]{FFFFFF}N/A} \\ \midrule
\rowcolor[HTML]{FFFFFF} 
Crutches &
   &
  \multicolumn{1}{c}{\cellcolor[HTML]{FFFFFF}\ding{51}} &
  \ding{51} &
  \ding{51} &
 Temporariness \\ \midrule
\rowcolor[HTML]{F3F3F3} 
White cane &
   &
  \multicolumn{1}{c}{\cellcolor[HTML]{F3F3F3}\ding{51}} &
  \ding{51} &
  \multicolumn{1}{l}{\cellcolor[HTML]{F3F3F3}} &
  \cellcolor[HTML]{FFFFFF}Visual study \\ \midrule
\rowcolor[HTML]{FFFFFF} 
Artificial limb or prosthetic &
   &
   &
  \ding{51} &
  \ding{51} &
  \cellcolor[HTML]{FFFFFF} \\
\rowcolor[HTML]{F3F3F3} 
Orthopaedic footwear &
   &
   &
  \multicolumn{1}{l}{\cellcolor[HTML]{F3F3F3}} &
  \ding{51} &
  \cellcolor[HTML]{FFFFFF} \\
\rowcolor[HTML]{FFFFFF} 
Orthotic or brace &
   &
   &
  \multicolumn{1}{l}{\cellcolor[HTML]{FFFFFF}} &
  \ding{51} &
  \multirow{-3}{*}{\cellcolor[HTML]{FFFFFF}Highly customized to the individual} \\ \midrule
\rowcolor[HTML]{F3F3F3} 
Sip and puff/tongue-controlled technology &
   &
   &
  \ding{51} &
  \multicolumn{1}{l}{\cellcolor[HTML]{F3F3F3}} &
  \cellcolor[HTML]{FFFFFF}Input device \\ \midrule
\rowcolor[HTML]{FFFFFF} 
Oxygen or breathing equipment &
   &
   &
  \ding{51} &
  \multicolumn{1}{l}{\cellcolor[HTML]{FFFFFF}} &
  \cellcolor[HTML]{FFFFFF} \\
\rowcolor[HTML]{F3F3F3} 
Augmented or alternative communication device &
   &
   &
  \ding{51} &
  \multicolumn{1}{l}{\cellcolor[HTML]{F3F3F3}} &
  \cellcolor[HTML]{FFFFFF} \\
\rowcolor[HTML]{FFFFFF} 
Grasping tool or reach extender &
   &
   &
  \multicolumn{1}{l}{\cellcolor[HTML]{FFFFFF}} &
  \ding{51} &
  \cellcolor[HTML]{FFFFFF} \\
\rowcolor[HTML]{F3F3F3} 
Adapted tools, utensils or special grips &
   &
   &
  \multicolumn{1}{l}{\cellcolor[HTML]{F3F3F3}} &
  \ding{51} &
  \cellcolor[HTML]{FFFFFF} \\
\rowcolor[HTML]{FFFFFF} 
Devices for dressing (\textit{e.g.}, button hook) &
   &
   &
  \multicolumn{1}{l}{\cellcolor[HTML]{FFFFFF}} &
  \ding{51} &
  \cellcolor[HTML]{FFFFFF} \\
\rowcolor[HTML]{F3F3F3} 
Device with oversized buttons (\textit{e.g.}, remote control or telephone) &
   &
   &
  \multicolumn{1}{l}{\cellcolor[HTML]{F3F3F3}} &
  \ding{51} &
  \multirow{-6}{4cm}{\cellcolor[HTML]{FFFFFF}Not exclusive to traveling outdoors using sidewalks} \\ \midrule
\rowcolor[HTML]{FFFFFF} 
Service animal &
   &
  \multicolumn{1}{c}{\cellcolor[HTML]{FFFFFF}\ding{51}} &
  \ding{51} &
  \multicolumn{1}{l}{\cellcolor[HTML]{FFFFFF}} &
  \cellcolor[HTML]{FFFFFF} \\
\rowcolor[HTML]{F3F3F3} 
Another person helping &
   &
   &
  \ding{51} &
  \multicolumn{1}{l}{\cellcolor[HTML]{F3F3F3}} &
  \multirow{-2}{*}{\cellcolor[HTML]{FFFFFF}Not a device} \\ \bottomrule
\end{tabular}%
}

\caption{
Our five mobility aid groups were informed by the response options in three surveys: the National Household Travel Survey (NHTS) question \textit{``Do you use any of the following?"}~\cite{us_department_of_transporation_national_2022}, the National Survey on Health and Disability (NSHD) question \textit{``When going outside your home and into the community what types of equipment, devices or help do you use?"}~\cite{the_university_of_kansas_national_2018}, and the Canadian Survey on Disability (CSD) question \textit{``Because of your condition, do you use any of the following?"}~\cite{government_of_canada_canadian_2022}. The table lists devices mentioned in the three surveys but excluded from our study's user groups, along with the rationale for each exclusion.}
\label{tab:survey-groups}
\end{table*}

\end{document}